\documentclass[a4paper, 11pt]{article}
\usepackage{float}
\usepackage{jheppub}
\usepackage{type1cm} % Allows the Computer Modern fonts at arbitrary sizes
\usepackage{graphicx} % Required for inserting images
\usepackage{tikz}
\usetikzlibrary{arrows.meta}
\usepackage{amsmath,amssymb,mathtools,bm}
\usepackage{physics}
\usepackage{hyperref}

\def\ol{\overline}

\hypersetup{
    colorlinks=true,
    urlcolor=blue,
    citecolor=blue,
    linkcolor=blue,
}

\let\originaltableofcontents\tableofcontents
\renewcommand\tableofcontents{%
    \begingroup
        \hypersetup{linktoc=page}%
        \originaltableofcontents
    \endgroup
}

\title{Spin-1 Bulk Local States in AdS/CFT and Carrollian Holography}
\author{\fontsize{10.4}{12.5}\selectfont Peng-Xiang Hao$^1$, Kotaro Shinmyo$^2$, Kenta Suzuki$^3$, Yu-ki Suzuki$^4$, Shunta Takahashi$^{5,6,7}$}
\affiliation[1]{Hetao Institute of Mathematics and Interdisciplinary Sciences, Futian Bonded Area, Shenzhen 518045, China}
\affiliation[2]{Center for Gravitational Physics and Quantum Information, Yukawa Institute for Theoretical Physics, Kyoto University, Kyoto 606-8502, Japan}
\affiliation[3]{Graduate School of Arts and Sciences, University of Tokyo, Komaba, Meguro-ku, Tokyo 153-8902, Japan}
\affiliation[4]{RIKEN Center for Interdisciplinary Theoretical and Mathematical Sciences (iTHEMS),\\ Wako, Saitama 351-0198, Japan}
\affiliation[5]{Perimeter Institute for Theoretical Physics, Waterloo, Ontario N2L 2Y5, Canada}
\affiliation[6]{Department of Physics and Astronomy, University of Waterloo, Waterloo, Ontario N2L 3G1, Canada}
\affiliation[7]{Department of Physics, Kyoto University, Kyoto 606-8502, Japan}

\date{}

\abstract{
    We construct the bulk local states of a massive spin-1 field in AdS$_3$/CFT$_2$ and three-dimensional Carrollian holography.
    In AdS, covariance under the isometries that leave a bulk point invariant determines the three components of the vector there and splits the solution into two branches of definite helicity, whose inner products reproduce the massive vector two-point function of AdS$_3$.
    By expanding the massive vector HKLL representation in boundary descendants, one can reproduce the three component states term by term.
    The two branches together describe the parity-invariant Proca field, and one alone the single propagating mode of Maxwell--Chern--Simons theory.
    The same conditions are solved in the induced representation of the Carrollian theory, where the solution again splits into two branches.
    Under the flat limit of the generators the highest weight module becomes the induced module, and the AdS states and Green's functions become their Carrollian counterparts, which supports the induced representation as the home of massive bulk excitations in Carrollian holography.
    This is the first bulk local state built with spin, complete from the state itself to its bulk-to-bulk propagator and to the limit relating the two descriptions.
}

\begin{document}
    \begin{flushright}
        YITP-26-135
        \\
        KUNS-3118
        \\
         RIKEN-iTHEMS-Report-26
    \end{flushright}

    \maketitle

    \section{Introduction}
        The holographic principle posits that gravitational physics in a region of spacetime can be described by degrees of freedom on its boundary \cite{tHooft:1993dmi,Susskind:1994vu}.
        The anti-de Sitter/conformal field theory (AdS/CFT) correspondence realizes this idea in a setting where the boundary theory supplies a microscopic description of bulk gravity \cite{Maldacena:1997re,Gubser:1998bc,Witten:1998qj}.
        At the level of quantum states, the aim is to identify the bulk and boundary Hilbert spaces, schematically
        \begin{align}
            \mathcal{H}_{\text{bulk}}\longleftrightarrow\mathcal{H}_{\text{boundary}}.
            \label{eq:hilbert-space-goal}
        \end{align}
        Bulk reconstruction asks how local fields and their dynamics emerge from that description.
        In AdS, the Hamilton-Kabat-Lifschytz-Lowe (HKLL) prescription reconstructs a bulk operator by smearing boundary operators \cite{Hamilton:2005ju,Hamilton:2006az}, with extensions to gauge fields, gravity, and massive vectors \cite{Kabat:2012hp}.

        A different approach is to reconstruct a local excitation at a point of the bulk from a superposition of boundary states whose coefficients encode the position of the bulk excitation \cite{Verlinde:2015qfa,Miyaji:2015fia,Nakayama:2015mva}.
        This is done by identifying the global conformal subalgebra of the Virasoro algebra on the boundary CFT with the bulk Killing vectors, and then translating the bulk symmetry conditions into constraint equations on a boundary state.
        The inner product of two such states returns the bulk two-point function, and this description is shown to be equivalent to the HKLL reconstruction in AdS/CFT \cite{Goto:2017olq}.
        To date, bulk local state reconstruction is restricted to scalar excitations but extended to Carrollian holography (flat holography) \cite{Hao:2025btl,Hao:2026cqm}, and dS/CFT \cite{Doi:2024nty}.
        Their inner product also provides access to bulk geometry through an information metric \cite{Miyaji:2015fia,Doi:2024nty, Hao:2025btl}.

        The remaining question is whether this explicit relation between local states, propagators, and boundary representations extends to spin.
        A vector is not invariant under the transformations that fix its position.
        Its three components mix, so their descendant coefficients must solve coupled covariance conditions.
        The scalar results therefore leave open how to construct the full triplet, recover its tensor two-point function, and relate its AdS and Carrollian representation labels.

        In this work, we construct massive spin-1 bulk local states in AdS$_3$/CFT$_2$ and three-dimensional Carrollian holography in the free one-particle sector.
        We solve the covariance conditions component by component, obtaining two branches with particle helicities $\lambda=\pm1$.
        In AdS, their overlaps reproduce the massive vector two-point function, and an independently normalized global mode expansion of the Proca field agrees with the descendant coefficients of the local states.
        This establishes their equivalence to HKLL reconstruction at this order.
        In the Carrollian theory, we construct the states in the induced representation, specify their algebraic duals, and evaluate the resulting pairings to obtain the Euclidean vector two-point function.
        The two helicities together describe the parity-invariant Proca field, whereas each branch has the local propagating content of Maxwell-Chern-Simons theory \cite{Deser:1981wh,Deser:1982vy}.
        If, instead, the highest weight representation (HWR) is selected, the local covariance and massive spin-1 Casimir conditions admit only the zero state, which suggests that the HWR is not suitable for reconstructing a non-trivial bulk local state.

        We also relate the AdS and Carrollian descriptions including spin at the level of their generators, primary labels, and descendant bases, as is done in ref.\,\cite{Hao:2025btl} for a scalar sector.
        The flat limit takes the AdS highest weight representation to the induced representation, with energy label $\xi=m$ and rotation weight $\Delta=\lambda$.
        An exact change of descendant basis at finite AdS radius makes this relation explicit.
        For regulated AdS local states with a fixed rotation component, we calculate the information metric and Berry curvature, which measure respectively the change in overlap magnitude and phase under displacement on a spatial slice.
        Their flat limit at fixed physical resolution retains a component-dependent metric while the Berry curvature vanishes.
        The massless analysis separately distinguishes the propagating Maxwell sector from Chern--Simons theory and examines the scalar descendants that survive the shortening of the massive representation.

        These constructions show how boundary descendants encode both the position and the vector character of a local excitation.
        They extend the scalar evidence for induced representations to massive spin one, with the tensor propagator and the AdS mode expansion providing tests beyond the symmetry algebra alone.
        They specify boundary states and pairings for localized one-particle excitations within the proposed correspondence in eq.\,\eqref{eq:hilbert-space-goal}, while leaving the completion of the Hilbert spaces and the interacting correspondence open.

\begin{figure}[h]
            \centering
            \begin{tikzpicture}[
                    every node/.style={draw, rectangle, align=center, minimum width=4.2cm, minimum height=1.1cm, font=\small},
                ]
                \node (massless) at (-4.5,1.75) {\textbf{Section~\ref{app:massless}}\\Massless vector case};
                \node (ads) at (0,3) {\textbf{Section~\ref{sec:ads}}\\AdS/CFT};
                \node (carr) at (5,3) {\textbf{Section~\ref{sec:intrinsic}}\\Carrollian holography};
                \node (flat) at (2.5,0.5) {\textbf{Section~\ref{sec:flat-limit}}\\Flat limit from AdS/CFT};
                \draw[->] (ads) -- (flat);
                \draw[->] (carr) -- (flat);
                \draw[->] (ads.south) |- (massless.east);
            \end{tikzpicture}
            \caption{Logical flow of the paper.
            Sections~\ref{sec:ads} and \ref{sec:intrinsic} are independent, while Section~\ref{sec:flat-limit} draws on both and Section~\ref{app:massless} specializes Section~\ref{sec:ads} alone.}
            \label{fig:roadmap}
        \end{figure}
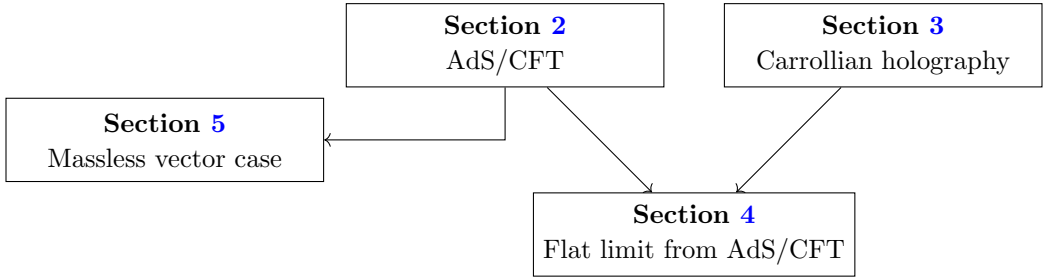

        This paper is organized as follows (see also Fig.\,\ref{fig:roadmap}).
        Section \ref{sec:ads} develops the AdS states, their propagators, the HKLL comparison,  the information metric and the Berry curvature.
        Section \ref{sec:intrinsic} gives the independent Carrollian construction and the highest weight obstruction.
        Section \ref{sec:flat-limit} relates the generators and descendant bases of these two descriptions.
        Section \ref{app:massless} specializes the AdS construction to the massless vector, and Section \ref{sec:conclusion} summarizes the results and discusses some future directions.
        Appendix \ref{appendix:ads-descendant-conversion} proves the change of descendant basis.

    \section{AdS/CFT}
        \label{sec:ads}

        We first construct the massive spin-1 local state in AdS$_3$/CFT$_2$.
        The construction extends the scalar twisted Ishibashi state by requiring the three helicity components to transform as a vector under the subgroup of the AdS isometry group that fixes the bulk point \cite{Miyaji:2015fia,Nakayama:2015mva}.
        We choose the relative phases of these components so that their flat limit agrees directly with the conventions of Section~\ref{sec:intrinsic}.

        \subsection{AdS spin-1 solution in the highest weight representation}

            Let $\ell$ be the AdS radius and $\Delta$ the lowest energy of the massive vector representation.
            The two helicity branches are generated from primary states with weights
            \begin{align}
                h_{\lambda}=\frac{\Delta+\lambda}{2},\qquad
                \bar h_{\lambda}=\frac{\Delta-\lambda}{2},\qquad
                \lambda=\pm1.
                \label{eq:ads-vector-weights}
            \end{align}
            The vector mass is related to the lowest energy by
            \begin{align}
                m^2\ell^2=(\Delta-1)^2,
                \label{eq:ads-vector-mass}
            \end{align}
            and we take the branch $\Delta=1+m\ell$ for $m>0$ \cite{Metsaev:2003cu}.
            We denote the corresponding highest weight states by $|h_{\lambda},\bar h_{\lambda}\rangle_{\lambda}$ and normalize their global descendants as
            \begin{align}
                |k,\bar k\rangle_{\lambda}
                =\frac{\mathcal{L}_{-1}^{k}\overline{\mathcal{L}}_{-1}^{\bar k}}
                {\sqrt{k!\,\bar k!\,(2h_{\lambda})_{k}(2\bar h_{\lambda})_{\bar k}}}
                |h_{\lambda},\bar h_{\lambda}\rangle_{\lambda}.
                \label{eq:ads-normalized-descendants}
            \end{align}
            Here $(x)_{k}:=\Gamma(x+k)/\Gamma(x)$ is the Pochhammer symbol.

            At the center of AdS, covariance requires the helicity components $|\Phi_{q}^{(\lambda)}\rangle$, with $q=1,0,-1$, to satisfy
            \begin{align}
                (\mathcal{L}_{0}-\overline{\mathcal{L}}_{0})|\Phi_{q}^{(\lambda)}\rangle
                &=q|\Phi_{q}^{(\lambda)}\rangle,
                \label{eq:ads-local-L0}\\
                (\mathcal{L}_{1}+\overline{\mathcal{L}}_{-1})|\Phi_{q}^{(\lambda)}\rangle
                &=\sqrt{2}\,|\Phi_{q-1}^{(\lambda)}\rangle,
                \label{eq:ads-local-L1}\\
                (\mathcal{L}_{-1}+\overline{\mathcal{L}}_{1})|\Phi_{q}^{(\lambda)}\rangle
                &=-\sqrt{2}\,|\Phi_{q+1}^{(\lambda)}\rangle,
                \label{eq:ads-local-L-1}
            \end{align}
            where $|\Phi_{\pm2}^{(\lambda)}\rangle=0$.
            These signs fix the phase convention for the helicity-zero component.
            They also make the contraction of eqs.\,\eqref{eq:ads-local-L0}--\eqref{eq:ads-local-L-1} identical to eqs.\,\eqref{eq:intrinsic-L0}--\eqref{eq:intrinsic-B}.

            For the $\lambda=+1$ branch, the solution is
            \begin{align}
                |\Phi_{1}^{(+)}\rangle
                &=\mathcal{N}_{A}e^{\frac{i\pi}{2}(\mathcal{L}_{0}+\overline{\mathcal{L}}_{0}-\Delta)}
                \sum_{k=0}^{\infty}a_{k}|k,k\rangle_{+},
                \label{eq:ads-state-plus}\\
                |\Phi_{0}^{(+)}\rangle
                &=i\sqrt{2}\,\mathcal{N}_{A}e^{\frac{i\pi}{2}(\mathcal{L}_{0}+\overline{\mathcal{L}}_{0}-\Delta)}
                \sum_{k=0}^{\infty}b_{k}|k,k+1\rangle_{+},
                \label{eq:ads-state-zero}\\
                |\Phi_{-1}^{(+)}\rangle
                &=-\mathcal{N}_{A}e^{\frac{i\pi}{2}(\mathcal{L}_{0}+\overline{\mathcal{L}}_{0}-\Delta)}
                \sum_{k=0}^{\infty}c_{k}|k,k+2\rangle_{+},
                \label{eq:ads-state-minus}
            \end{align}
            with
            \begin{align}
                a_{k}&=\sqrt{\frac{(\Delta+1)_{k}}{(\Delta-1)_{k}}},
                \label{eq:ads-ak}\\
                b_{k}&=a_{k}\sqrt{\frac{k+1}{\Delta+k-1}},
                \label{eq:ads-bk}\\
                c_{k}&=a_{k}\sqrt{\frac{(k+1)(k+2)}{(\Delta+k-1)(\Delta+k)}}.
                \label{eq:ads-ck}
            \end{align}
            The action of the global conformal generators on the normalized descendants verifies eqs.\,\eqref{eq:ads-local-L0}--\eqref{eq:ads-local-L-1} term by term.
            In particular, the square roots in eqs.\,\eqref{eq:ads-bk} and \eqref{eq:ads-ck} follow from the norms of the off-diagonal descendants.

            The $\lambda=-1$ branch follows by interchanging the holomorphic and antiholomorphic modules:
            \begin{align}
                |\Phi_{-1}^{(-)}\rangle
                &=\mathcal{N}_{A}e^{\frac{i\pi}{2}(\mathcal{L}_{0}+\overline{\mathcal{L}}_{0}-\Delta)}
                \sum_{k=0}^{\infty}a_{k}|k,k\rangle_{-},
                \label{eq:ads-state-negative-minus}\\
                |\Phi_{0}^{(-)}\rangle
                &=-i\sqrt{2}\,\mathcal{N}_{A}e^{\frac{i\pi}{2}(\mathcal{L}_{0}+\overline{\mathcal{L}}_{0}-\Delta)}
                \sum_{k=0}^{\infty}b_{k}|k+1,k\rangle_{-},
                \label{eq:ads-state-negative-zero}\\
                |\Phi_{1}^{(-)}\rangle
                &=-\mathcal{N}_{A}e^{\frac{i\pi}{2}(\mathcal{L}_{0}+\overline{\mathcal{L}}_{0}-\Delta)}
                \sum_{k=0}^{\infty}c_{k}|k+2,k\rangle_{-}.
                \label{eq:ads-state-negative-plus}
            \end{align}
            A parity-invariant Proca field contains both branches, whereas a single-helicity massive vector contains only one of them.
            The state at any other bulk point is obtained from eqs.\,\eqref{eq:ads-state-plus}--\eqref{eq:ads-state-negative-plus} by the AdS isometry that maps the center to that point.

        \subsection{Bulk propagators}

            The inner product of two bulk local states gives their bulk two-point function \cite{Miyaji:2015fia,Nakayama:2015mva}.
            AdS covariance reduces the calculation at spacelike separation to two points on a common geodesic.
            If $\chi=D(X,X')/\ell$ is their dimensionless geodesic distance, the helicity components of the positive branch can be evaluated at the center as
            \begin{align}
                G_{11}^{(+)}(\chi)
                &=|\mathcal{N}_{A}|^2e^{-\Delta\chi}
                \sum_{k=0}^{\infty}a_{k}^{2}e^{-2k\chi},
                \label{eq:ads-Gpp-sum}\\
                G_{00}^{(+)}(\chi)
                &=2|\mathcal{N}_{A}|^2e^{-(\Delta+1)\chi}
                \sum_{k=0}^{\infty}b_{k}^{2}e^{-2k\chi},
                \label{eq:ads-G00-sum}\\
                G_{-1-1}^{(+)}(\chi)
                &=|\mathcal{N}_{A}|^2e^{-(\Delta+2)\chi}
                \sum_{k=0}^{\infty}c_{k}^{2}e^{-2k\chi}.
                \label{eq:ads-Gmm-sum}
            \end{align}
            Components with different helicities vanish in the frame parallel transported along the geodesic.
            The descendant coefficients obey
            \begin{align}
                a_{k}^{2}&=\frac{(\Delta+k)(\Delta+k-1)}{\Delta(\Delta-1)},\\
                b_{k}^{2}&=\frac{(k+1)(\Delta+k)}{\Delta(\Delta-1)},\\
                c_{k}^{2}&=\frac{(k+1)(k+2)}{\Delta(\Delta-1)}.
                \label{eq:ads-coefficient-squares}
            \end{align}
            Therefore the sums in eqs.\,\eqref{eq:ads-Gpp-sum}--\eqref{eq:ads-Gmm-sum} reduce to derivatives of a geometric series and give
            \begin{align}
                G_{11}^{(+)}(\chi)
                &=\frac{|\mathcal{N}_{A}|^2e^{-(\Delta-1)\chi}}
                {4\Delta(\Delta-1)\sinh^3\!\chi}
                \big(1+2(\Delta-1)\sinh\!\chi\cosh\!\chi
                +2(\Delta-1)^2\sinh^2\!\chi\big),
                \label{eq:ads-Gpp}\\
                G_{00}^{(+)}(\chi)
                &=\frac{2|\mathcal{N}_{A}|^2e^{-(\Delta-1)\chi}}
                {4\Delta(\Delta-1)\sinh^3\!\chi}
                \big(\cosh\!\chi+(\Delta-1)\sinh\!\chi\big),
                \label{eq:ads-G00}\\
                G_{-1-1}^{(+)}(\chi)
                &=\frac{|\mathcal{N}_{A}|^2e^{-(\Delta-1)\chi}}
                {4\Delta(\Delta-1)\sinh^3\!\chi}.
                \label{eq:ads-Gmm}
            \end{align}
            The factor of two in eq.\,\eqref{eq:ads-G00} is fixed by the normalized spin-1 basis used in eqs.\,\eqref{eq:ads-local-L1} and \eqref{eq:ads-local-L-1}.
            For the negative branch, $G_{11}^{(-)}=G_{-1-1}^{(+)}$, $G_{00}^{(-)}=G_{00}^{(+)}$, and $G_{-1-1}^{(-)}=G_{11}^{(+)}$.

            Parallel transport of the helicity frame reconstructs the two-point function for arbitrary $X$ and $X'$.
            The result has the maximally symmetric vector-bitensor form and agrees, up to the overall choice of $\mathcal{N}_{A}$, with the massive vector two-point function in AdS$_3$ \cite{Allen:1985wd}.
            The sum of the two helicity branches gives the parity-even Proca propagator, while their difference isolates the parity-odd helicity dependence.
            Lorentzian Wightman, Feynman, retarded, and advanced propagators follow from the corresponding analytic continuation and $i\epsilon$ prescription.

        \subsection{Relation to HKLL reconstruction}
            \label{subsec:ads-hkll}

            The HKLL comparison admits an entirely global derivation.
            In the free theory, the HKLL operator is the regular normalizable solution whose boundary limit equals the prescribed CFT operator; expanding that solution in global normal modes is therefore a global HKLL reconstruction \cite{Hamilton:2006az,Kabat:2012hp}.
            At the center of AdS, the values of those modes determine the action of the reconstructed operator on the vacuum.
            This is the spin-1 counterpart of the scalar mode comparison in Ref.\,\cite{Goto:2017olq}, but the Proca equation and the canonical normalization of its modes replace the scalar radial equation.
            The calculation below concerns the one-particle sector at leading order in the semiclassical expansion.

            \subsubsection{Global Proca modes and the HKLL operator}
                \label{subsubsec:ads-global-hkll}

                We set $\ell=1$ in this subsection and write
                \begin{align}
                    ds^2=-(1+r^2)d\tau^2+\frac{dr^2}{1+r^2}+r^2d\theta^2,
                    \qquad
                    M:=\Delta-1.
                    \label{eq:ads-hkll-global-metric}
                \end{align}
                A single helicity of the Proca field obeys the first-order equation
                \begin{align}
                    \epsilon^{MNP}\nabla_N A_P=-M A^M.
                    \label{eq:ads-hkll-self-dual-equation}
                \end{align}
                It implies both $\nabla_MA^M=0$ and $(\nabla^2-M^2+2)A_M=0$; changing the sign on the right-hand side selects the opposite helicity.
                The two signs together describe the parity-invariant Proca field, while either sign may be quantized separately \cite{Bhat:2025ads}.

                The combinations $A_{\pm}=A_\tau\pm A_\theta$ decouple under the scalar Laplacian $\Box$,
                \begin{align}
                    \bigl(\Box-(M+1)^2+1\bigr)A_+&=0,
                    &
                    \bigl(\Box-(M-1)^2+1\bigr)A_-&=0.
                    \label{eq:ads-hkll-decoupled-components}
                \end{align}
                Regularity at $r=0$ and normalizability at the boundary restrict the frequencies to
                \begin{align}
                    \omega_{p,n}=M+|p|+2n,
                    \qquad
                    \begin{cases}
                        n=0, & p=1,2,\ldots,\\
                        n=1,2,\ldots, & p\in\mathbb{Z}.
                    \end{cases}
                    \label{eq:ads-hkll-global-spectrum}
                \end{align}
                For $r=\tan\sigma$, the radial functions of a mode proportional to $e^{-i\omega_{p,n}\tau+ip\theta}$ may be written as \cite{Bhat:2025ads}
                \begin{align}
                    R_-^{(p,n)}(\sigma)
                    &=C_-^{(p,n)}(-1)^n\frac{n!\,|p|!}{(n+|p|)!}
                    (\cos\sigma)^M(\sin\sigma)^{|p|}
                    P_n^{(M-1,|p|)}(-\cos2\sigma),
                    \nonumber\\
                    R_+^{(p,n)}(\sigma)
                    &=C_+^{(p,n)}(-1)^{n-1}
                    \frac{(n-1)!\,|p|!}{(n-1+|p|)!}
                    (\cos\sigma)^{M+2}(\sin\sigma)^{|p|}
                    P_{n-1}^{(M+1,|p|)}(-\cos2\sigma).
                    \label{eq:ads-hkll-global-radial-modes}
                \end{align}
                The second line is absent when $n=0$.
                The endpoint values needed at the boundary and at the center are
                \begin{align}
                    P_n^{(\alpha,\beta)}(-1)
                    &=(-1)^n\frac{(\beta+1)_n}{n!},
                    &
                    P_n^{(\alpha,\beta)}(1)
                    &=\frac{(\alpha+1)_n}{n!}.
                    \label{eq:ads-hkll-jacobi-endpoints}
                \end{align}
                The first-order equation fixes the remaining constants and the radial component:
                \begin{align}
                    \frac{C_+^{(p,n)}}{C_-^{(p,n)}}
                    &=\frac{M+|p|-\operatorname{sgn}(p)\omega_{p,n}}
                    {M-|p|-\operatorname{sgn}(p)\omega_{p,n}},
                    &&p\neq0,
                    \nonumber\\
                    C_+^{(0,n)}&=C_-^{(0,n)},
                    \nonumber\\
                    R_r^{(p,n)}
                    &=-\frac{i}{2M(1+r^2)r}
                    \Big((p+\omega_{p,n})R_+^{(p,n)}
                    +(p-\omega_{p,n})R_-^{(p,n)}\Big).
                    \label{eq:ads-hkll-mode-constraints}
                \end{align}
                These regular normalizable modes form a complete basis of the one-particle solution space.
                Canonical normalization fixes $C_-^{(p,n)}$ for each mode, and eq.\,\eqref{eq:ads-hkll-mode-constraints} then fixes all three vector components.
                Writing the resulting vector-valued mode as $u_M^{(p,n)}(X)$, the field expansion is
                \begin{align}
                    A_M(X)=\sum_{p,n}
                    \left[u_M^{(p,n)}(X)a_{p,n}
                    +u_M^{(p,n)*}(X)a_{p,n}^{\dagger}\right],
                    \qquad
                    [a_{p,n},a_{p',n'}^{\dagger}]
                    =\delta_{pp'}\delta_{nn'}.
                    \label{eq:ads-hkll-global-mode-expansion}
                \end{align}

                The extrapolate dictionary turns this mode expansion into the global HKLL map.
                It identifies the boundary primary of weights
                $(h,\bar{h})=((\Delta+1)/2,(\Delta-1)/2)$ with
                \begin{align}
                    J(\tau,\theta)
                    =\lim_{r\to\infty}r^M A_-(\tau,r,\theta),
                    \label{eq:ads-hkll-global-extrapolate}
                \end{align}
                up to the overall normalization of $J$.
                The boundary coefficient follows from the second Jacobi endpoint in eq.\,\eqref{eq:ads-hkll-jacobi-endpoints}:
                \begin{align}
                    R_-^{(p,n)}(r)&\sim B_{p,n}r^{-M},
                    &
                    B_{p,n}
                    &=(-1)^n C_-^{(p,n)}
                    \frac{|p|!\,(M)_n}{(n+|p|)!}.
                    \label{eq:ads-hkll-boundary-coefficients}
                \end{align}
                None of the coefficients $B_{p,n}$ vanishes for $M>0$, so the boundary modes determine every oscillator in eq.\,\eqref{eq:ads-hkll-global-mode-expansion}.
                The cylinder expansion of $J$ consequently takes the form
                \begin{align}
                    J(\tau,\theta)
                    =\sum_{p,n}\left[
                    J_{p,n}e^{-i\omega_{p,n}\tau+ip\theta}
                    +J_{p,n}^{\dagger}e^{i\omega_{p,n}\tau-ip\theta}
                    \right],
                    \qquad
                    J_{p,n}=B_{p,n}a_{p,n}.
                    \label{eq:ads-hkll-boundary-modes}
                \end{align}
                The resulting global HKLL reconstruction is
                \begin{align}
                    A_M^{\mathrm{HKLL}}(X)
                    =\sum_{p,n}\left[
                    \frac{u_M^{(p,n)}(X)}{B_{p,n}}J_{p,n}
                    +\frac{u_M^{(p,n)*}(X)}{B_{p,n}^*}J_{p,n}^{\dagger}
                    \right].
                    \label{eq:ads-global-vector-hkll}
                \end{align}
                Fourier inversion converts eq.\,\eqref{eq:ads-global-vector-hkll} into the usual boundary smearing integral.
                We retain the expression in mode space because each term maps directly to one global descendant.

            \subsubsection{The reconstructed state at the center}
                \label{subsubsec:ads-hkll-state-consistency}

                The center at $\tau=0$ is denoted by $X_0$.
                The orthonormal polar components are
                $A_{\hat{0}}=A_\tau/\sqrt{1+r^2}$,
                $A_{\hat{r}}=\sqrt{1+r^2}A_r$, and
                $A_{\hat{\theta}}=A_\theta/r$.
                Their circular combinations in a Cartesian frame are
                \begin{align}
                    \widehat{A}_{\pm1}
                    &=\pm\frac{i}{\sqrt{2}}e^{\pm i\theta}
                    (A_{\hat{r}}\pm iA_{\hat{\theta}}),
                    &
                    \widehat{A}_0&=A_{\hat{0}}.
                    \label{eq:ads-hkll-circular-components}
                \end{align}
                The phases in eq.\,\eqref{eq:ads-hkll-circular-components} implement the convention of eqs.\,\eqref{eq:ads-local-L0}--\eqref{eq:ads-local-L-1}.
                The factors $e^{\pm i\theta}$ ensure that only modes with $p=q$ remain at $r=0$.
                The three surviving towers are
                \begin{align}
                    (p,n)=(1,k),\qquad (0,k+1),\qquad (-1,k+1),
                    \qquad k=0,1,2,\ldots.
                    \label{eq:ads-hkll-center-mode-towers}
                \end{align}

                The first Jacobi endpoint in eq.\,\eqref{eq:ads-hkll-jacobi-endpoints} fixes the center values and their phases relative to the boundary data.
                The coefficient $B_{p,n}$ in eq.\,\eqref{eq:ads-hkll-boundary-coefficients} contains the factor $(-1)^n$, whereas the leading coefficient at the center is positive for $C_-^{(p,n)}>0$.
                The canonical normalization constants for the three towers in eq.\,\eqref{eq:ads-hkll-center-mode-towers} are
                \begin{align}
                    C_-^{(1,k)}
                    &=(k+1)\sqrt{\frac{(M+k)(M+k+1)}{\pi M}},
                    \nonumber\\
                    C_-^{(0,k+1)}
                    &=\sqrt{\frac{(k+1)(M+k+1)}{\pi M}},
                    \nonumber\\
                    C_-^{(-1,k+1)}
                    &=(M+k+1)\sqrt{\frac{(k+1)(k+2)}{\pi M}}.
                    \label{eq:ads-hkll-relevant-mode-normalizations}
                \end{align}
                The corresponding center values of the circular components are respectively
                $C_-^{(1,k)}/[\sqrt{2}(k+1)]$,
                $C_-^{(0,k+1)}$, and
                $-C_-^{(-1,k+1)}/[\sqrt{2}(M+k+1)]$.
                The last sign is the relative sign between the two spatial circular components in the phase convention of eqs.\,\eqref{eq:ads-local-L1} and \eqref{eq:ads-local-L-1}.
                With the canonical normalization in eq.\,\eqref{eq:ads-hkll-global-mode-expansion}, their values at $X_0$ are
                \begin{align}
                    \widehat{A}_1^{(+),\mathrm{HKLL}}(X_0)|0\rangle
                    &=\mathcal{C}\sum_{k=0}^{\infty}a_k
                    a_{1,k}^{\dagger}|0\rangle,
                    \nonumber\\
                    \widehat{A}_0^{(+),\mathrm{HKLL}}(X_0)|0\rangle
                    &=\sqrt{2}\,\mathcal{C}\sum_{k=0}^{\infty}b_k
                    a_{0,k+1}^{\dagger}|0\rangle,
                    \nonumber\\
                    \widehat{A}_{-1}^{(+),\mathrm{HKLL}}(X_0)|0\rangle
                    &=-\mathcal{C}\sum_{k=0}^{\infty}c_k
                    a_{-1,k+1}^{\dagger}|0\rangle,
                    \qquad
                    \mathcal{C}=\sqrt{\frac{\Delta}{2\pi}}.
                    \label{eq:ads-hkll-center-oscillators}
                \end{align}
                The value of $\mathcal{C}$ corresponds to the normalization of the bulk field used in Ref.\,\cite{Bhat:2025ads}; changing the normalization of $A_M$ rescales all three lines together.
                The common field normalization cancels from the three relative coefficients:
                \begin{align}
                    a_k^2&=\frac{(M+k)(M+k+1)}{M(M+1)},
                    &
                    b_k^2&=\frac{(k+1)(M+k+1)}{M(M+1)},
                    &
                    c_k^2&=\frac{(k+1)(k+2)}{M(M+1)}.
                    \label{eq:ads-hkll-center-coefficients}
                \end{align}
                With $M=\Delta-1$, they coincide with eqs.\,\eqref{eq:ads-ak}--\eqref{eq:ads-ck}.
                Here they follow from the radial Proca problem and its canonical inner product, not from the covariance conditions defining the bulk local states.

                The global energy and angular momentum of $a_{p,n}^{\dagger}|0\rangle$ determine its CFT descendant levels $(n_1,n_2)$:
                \begin{align}
                    2n+|p|&=n_1+n_2+1,
                    &
                    p&=n_1-n_2+1.
                    \label{eq:ads-hkll-mode-descendant-levels}
                \end{align}
                The correspondence between global modes and CFT descendants identifies the corresponding canonically normalized one-particle states with normalized descendants \cite{Bhat:2025ads}.
                The phases follow from the boundary coefficients $B_{p,n}$ fixed above:
                \begin{align}
                    a_{1,k}^{\dagger}|0\rangle
                    &=(-1)^k|k,k\rangle_+,
                    \nonumber\\
                    a_{0,k+1}^{\dagger}|0\rangle
                    &=(-1)^{k+1}|k,k+1\rangle_+,
                    \nonumber\\
                    a_{-1,k+1}^{\dagger}|0\rangle
                    &=(-1)^{k+1}|k,k+2\rangle_+.
                    \label{eq:ads-hkll-global-state-dictionary}
                \end{align}
                After division by the common factor $\mathcal{C}$, eq.\,\eqref{eq:ads-hkll-center-oscillators} becomes
                \begin{align}
                    \mathcal{A}_1^{(+),\mathrm{HKLL}}(X_0)|0\rangle
                    &=\sum_{k=0}^{\infty}(-1)^k a_k|k,k\rangle_+,
                    \nonumber\\
                    \mathcal{A}_0^{(+),\mathrm{HKLL}}(X_0)|0\rangle
                    &=-\sqrt{2}\sum_{k=0}^{\infty}(-1)^k b_k|k,k+1\rangle_+,
                    \nonumber\\
                    \mathcal{A}_{-1}^{(+),\mathrm{HKLL}}(X_0)|0\rangle
                    &=\sum_{k=0}^{\infty}(-1)^k c_k|k,k+2\rangle_+,
                    \qquad
                    \mathcal{A}_q^{\mathrm{HKLL}}:=\mathcal{C}^{-1}\widehat{A}_q^{\mathrm{HKLL}}.
                    \label{eq:ads-hkll-normalized-series}
                \end{align}
                The division by $\mathcal{C}$ only sets the coefficient of the primary in the first line to one, which is the convention used for the bulk local states; it does not alter any relative component or descendant coefficient.

                In the unnormalized descendant basis, the three lines of eq.\,\eqref{eq:ads-hkll-normalized-series} become
                \begin{align}
                    \mathcal{A}_1^{(+),\mathrm{HKLL}}(X_0)|0\rangle
                    &=\sum_{k=0}^{\infty}\frac{(-1)^k\mathcal{L}_{-1}^{k}\overline{\mathcal{L}}_{-1}^{k}}{k!(\Delta-1)_k}|h_+,\bar{h}_+\rangle,
                    \nonumber\\
                    \mathcal{A}_0^{(+),\mathrm{HKLL}}(X_0)|0\rangle
                    &=-\sqrt{2}\sum_{k=0}^{\infty}\frac{(-1)^k\mathcal{L}_{-1}^{k}\overline{\mathcal{L}}_{-1}^{k+1}}{k!(\Delta-1)_{k+1}}|h_+,\bar{h}_+\rangle,
                    \nonumber\\
                    \mathcal{A}_{-1}^{(+),\mathrm{HKLL}}(X_0)|0\rangle
                    &=\sum_{k=0}^{\infty}\frac{(-1)^k\mathcal{L}_{-1}^{k}\overline{\mathcal{L}}_{-1}^{k+2}}{k!(\Delta-1)_{k+2}}|h_+,\bar{h}_+\rangle.
                    \label{eq:ads-hkll-global-descendants}
                \end{align}
                Thus the global HKLL reconstruction produces the coefficients $a_k$, $b_k$, and $c_k$ term by term.

                The last ingredient in the comparison is the twist phase.
                The twist in eqs.\,\eqref{eq:ads-state-plus}--\eqref{eq:ads-state-minus} acts as
                \begin{align}
                    e^{\frac{i\pi}{2}(\mathcal{L}_0+\overline{\mathcal{L}}_0-\Delta)}
                    |k,k+s\rangle_+
                    =i^{2k+s}|k,k+s\rangle_+.
                    \label{eq:ads-global-hkll-twist-action}
                \end{align}
                For $s=0,1,2$, this action converts eq.\,\eqref{eq:ads-hkll-normalized-series} into eqs.\,\eqref{eq:ads-state-plus}--\eqref{eq:ads-state-minus}, including the relative factor and phase of the $q=0$ component.
                Hence
                \begin{align}
                    |\Phi_q^{(+)}(X_0)\rangle
                    =\mathcal{N}_A
                    \mathcal{A}_q^{(+),\mathrm{HKLL}}(X_0)|0\rangle,
                    \qquad q=1,0,-1.
                    \label{eq:ads-global-hkll-state-equivalence}
                \end{align}

                Reversing the sign in eq.\,\eqref{eq:ads-hkll-self-dual-equation} interchanges the holomorphic and antiholomorphic sectors and reverses the orientation of the circular frame.
                At $X_0$, the frame transformation is
                \begin{align}
                    \widetilde{\mathcal{A}}_{\pm1}
                    &=-\mathcal{A}_{\mp1},
                    &
                    \widetilde{\mathcal{A}}_0&=\mathcal{A}_0.
                    \label{eq:ads-hkll-opposite-helicity-frame}
                \end{align}
                Exchanging the holomorphic and antiholomorphic sectors leaves the symmetric coefficient triple $(1,-\sqrt{2},1)$ unchanged, and eq.\,\eqref{eq:ads-hkll-opposite-helicity-frame} then sends it to $(-1,-\sqrt{2},-1)$.
                Absorbing the common minus sign into the phase of the negative-helicity primary leaves the second branch in the form
                \begin{align}
                    \mathcal{A}_{-1}^{(-),\mathrm{HKLL}}(X_0)|0\rangle
                    &=\sum_{k=0}^{\infty}\frac{(-1)^k\mathcal{L}_{-1}^{k}\overline{\mathcal{L}}_{-1}^{k}}{k!(\Delta-1)_k}|h_-,\bar{h}_-\rangle,
                    \nonumber\\
                    \mathcal{A}_0^{(-),\mathrm{HKLL}}(X_0)|0\rangle
                    &=\sqrt{2}\sum_{k=0}^{\infty}\frac{(-1)^k\mathcal{L}_{-1}^{k+1}\overline{\mathcal{L}}_{-1}^{k}}{k!(\Delta-1)_{k+1}}|h_-,\bar{h}_-\rangle,
                    \nonumber\\
                    \mathcal{A}_1^{(-),\mathrm{HKLL}}(X_0)|0\rangle
                    &=\sum_{k=0}^{\infty}\frac{(-1)^k\mathcal{L}_{-1}^{k+2}\overline{\mathcal{L}}_{-1}^{k}}{k!(\Delta-1)_{k+2}}|h_-,\bar{h}_-\rangle.
                    \label{eq:ads-global-hkll-negative-descendants}
                \end{align}
                After the same twist, the normalized version of eq.\,\eqref{eq:ads-global-hkll-negative-descendants} coincides with eqs.\,\eqref{eq:ads-state-negative-minus}--\eqref{eq:ads-state-negative-plus}.
                AdS isometries extend eq.\,\eqref{eq:ads-global-hkll-state-equivalence} from $X_0$ to every bulk point.
                Because the calculation began with the independently normalized global solutions of the Proca equation, the agreement is a mode-by-mode check of the bulk local states rather than a consequence of their defining covariance conditions.

\subsection{Information metric and Berry curvature of spin-1 local states}
\label{sec:spinone-info-geometry}

The overlap of nearby normalized bulk local states defines a Bures information metric, following the scalar case \cite{Miyaji:2015fia}. For spin one, this metric also depends on the chosen vector component, while the phase of the overlap determines a Berry connection, which is a new ingredient in the spin one case. We calculate both quantities directly from the inner product for a specified regularization and study their flat limit. We restore the explicit AdS radius $\ell$ after setting $\ell=1$ in the preceding HKLL calculation. The particle helicity $\lambda=\pm1$ labels a representation, whereas $q=1,0,-1$ labels its local vector components.

\subsubsection{Information metric}
\label{sec:spinone-info-orbit}

Let
\begin{equation}
 H=\mathcal{L}_0+\overline{\mathcal{L}}_0,\qquad
 J=\mathcal{L}_0-\overline{\mathcal{L}}_0,
 \label{eq:spinone-info-HJ}
\end{equation}
and first consider the $\lambda=+1$ branch. Since the local states have divergent norms, we define the normalized states at the center by
\begin{equation}
 |v_q\rangle=
 \frac{e^{-\varepsilon H}|\Phi_q^{(+)}\rangle}
 {\bigl(\langle\Phi_q^{(+)}|e^{-2\varepsilon H}
                        |\Phi_q^{(+)}\rangle\bigr)^{1/2}},
 \qquad \varepsilon>0.
 \label{eq:spinone-info-seed}
\end{equation}
The overall normalization $\mathcal{N}_A$ cancels. Moreover, $[H,J]=0$ implies $J|v_q\rangle=q|v_q\rangle$.

The Hermitian generators
\begin{align}
 K_1&=\frac{\mathcal{L}_1+\mathcal{L}_{-1}
                  -\overline{\mathcal{L}}_1-\overline{\mathcal{L}}_{-1}}2,
 \nonumber\\
 K_2&=\frac{\mathcal{L}_{-1}-\mathcal{L}_1
                  +\overline{\mathcal{L}}_{-1}-\overline{\mathcal{L}}_1}{2i},
 \label{eq:spinone-info-generators}
\end{align}
satisfy
\begin{equation}
 [J,K_1]=iK_2,\qquad [J,K_2]=-iK_1,\qquad [K_1,K_2]=-iJ.
 \label{eq:spinone-info-algebra}
\end{equation}
Together with $J$, they generate the $SO(2,1)$ isometries of the constant-time hyperbolic slice.  We transport the regulated states by
\begin{equation}
 |\psi_q(\rho,\phi)\rangle=U(\rho,\phi)|v_q\rangle,
 \qquad U=e^{-i\phi J}e^{-i\rho K_2},
 \label{eq:spinone-info-orbit}
\end{equation}
where $r=\sinh \rho$ in the radius-one coordinates of eq.\,\eqref{eq:ads-hkll-global-metric}, $\phi$ is the angular coordinate, and the physical spatial metric is
$\ell^2(d\rho^2+\sinh^2\rho\,d\phi^2)$.

For the normalized pure states in eq.\,\eqref{eq:spinone-info-orbit}, let
\begin{equation}
 \mathcal S_q(x,x+\delta x)
 =\langle\psi_q(x)|\psi_q(x+\delta x)\rangle,
 \qquad x=(\rho,\phi).
 \label{eq:spinone-info-overlap}
\end{equation}
We adopt the overlap normalization of the information metric in Ref.\,\cite{Miyaji:2015fia}:
\begin{equation}
 1-|\mathcal S_q(x,x+\delta x)|
 =g^{\mathrm B,q}_{ij}(x)\,\delta x^i\delta x^j
      +O(|\delta x|^3),
 \qquad
 ds_{\mathrm B,q}^2
 =g^{\mathrm B,q}_{ij}\,dx^idx^j.
 \label{eq:spinone-info-bures-definition}
\end{equation}
Our normalization is one half of the commonly used squared Bures distance
$D_{\mathrm B}^2=2(1-|\mathcal S_q|)$ for pure states.
Using that convention instead multiplies every metric below by two.

All expectation values in the following calculation are taken in $|v_q\rangle$, where $\langle
\mathcal{O}\rangle:=\langle v_q|\mathcal{O}|v_q\rangle$. Rotational covariance and the definite $J$ eigenvalue give
\begin{equation}
 \langle K_1\rangle=\langle K_2\rangle=0,\qquad
 \langle K_1^2\rangle=\langle K_2^2\rangle:= V_q,
 \qquad \langle\{K_1,K_2\}\rangle=0,
 \label{eq:spinone-info-variance}
\end{equation}
where the explicit form of $V_q$ is given later.
The last commutator in eq.\,\eqref{eq:spinone-info-algebra} then fixes
$\langle K_1K_2\rangle=-iq/2$.
Using the commutation relations, we find
\begin{equation}
 e^{i\rho K_2}Je^{-i\rho K_2}=J\cosh\rho-K_1\sinh\rho.
 \label{eq:spinone-info-adjoint}
\end{equation}
The relative isometry between the two points therefore gives the exact overlap
\begin{equation}
 \mathcal S_q
 =\langle v_q|
 e^{-i\delta\phi(J\cosh\rho-K_1\sinh\rho)}
 e^{-i\delta\rho K_2}|v_q\rangle.
 \label{eq:spinone-info-relative-isometry}
\end{equation}
Expanding the two exponentials to second order and using $J|v_q\rangle=q|v_q\rangle$ and eq.\,\eqref{eq:spinone-info-variance}, we find
\begin{align}
 \mathcal S_q={}&1-iq\cosh\rho\,\delta\phi
 -\frac{V_q}{2}\delta\rho^2
 -\frac{V_q\sinh^2\rho+q^2\cosh^2\rho}{2}\delta\phi^2
 \nonumber\\
 &-\frac{iq}{2}\sinh\rho\,\delta\rho\,\delta\phi
 +O(|\delta x|^3).
 \label{eq:spinone-info-overlap-expansion}
\end{align}
The mixed term follows from
$\langle(J\cosh\rho-K_1\sinh\rho)K_2\rangle
=iq\sinh\rho/2$.
By direct computation, we obtain
\begin{align}
 |\mathcal S_q|
 =1-\frac{V_q}{2}(\delta\rho^2+\sinh^2\rho\,\delta\phi^2)
       +O(|\delta x|^3).
 \label{eq:spinone-info-overlap-modulus}
\end{align}
Thus, the Bures information metric in our convention is
\begin{equation}
 ds_{\mathrm B,q}^2
   =\mathcal G_q(\Delta,\varepsilon)
      (d\rho^2+\sinh^2\rho\,d\phi^2),
 \qquad \mathcal G_q=\frac{V_q}{2}.
 \label{eq:spinone-info-metric}
\end{equation}
The hyperbolic form holds for every finite regulator value. Its coefficient, calculated below, need not be the same for the three components.

The phase of the same overlap determines the Berry connection. With the convention $\mathcal{A}=i\langle\psi|d\psi\rangle$, the linear term in eq.\,\eqref{eq:spinone-info-overlap-expansion} gives $\mathcal{A}=q\cosh\rho\,d\phi$. Since the state has the following nontrivial angular dependence in the limit to the center
\begin{equation}
    \lim_{\rho\to 0}|\psi_q(\rho,\phi)\rangle=e^{-i\phi q}|v_q\rangle,
\end{equation}
the phase redefinition $|\psi\rangle\mapsto e^{iq\phi}|\psi\rangle$ makes the state regular at the origin. The connection then becomes
\begin{equation}
 \mathcal{A}_q=q(\cosh\rho-1)d\phi,
 \qquad \mathcal{F}_q=d\mathcal{A}_q
     =q\sinh\rho\,d\rho\wedge d\phi.
 \label{eq:spinone-info-berry}
\end{equation}

Since $U(\rho,\phi)$ is an isometry, we expect the Levi-Civita connection to be related to the Berry connection.
In the oriented orthonormal frame
\begin{equation}
    e^1=\ell d\rho, \quad e^2=\ell\sinh\rho\,d\phi,
    \end{equation}
Cartan's first structure equation is
\begin{equation}
    de^a+\omega^a{}_b\wedge e^b=0,
\end{equation}
where $\omega^a{}_b$ is the Levi-Civita connection. Using the antisymmetry of $\omega$, we write the equation in components as
\begin{equation}
    de^1+\omega^1{}_2\wedge e^2=0,\quad
    de^2+\omega^2{}_1\wedge e^1=0.
\end{equation}
The solution is
\begin{equation}
    \omega^2{}_1=\cosh\rho d\phi.
\end{equation}
The Berry connection is therefore proportional to the Levi-Civita connection with $\mathcal{A}=q\,\omega^2{}_1$ up to a gauge transformation. For a closed curve $C$ enclosing signed physical area $\operatorname{Area}(C)$,
\begin{equation}
 \gamma_q(C)=\oint_C \mathcal{A}_q=\frac{q}{\ell^2}\operatorname{Area}(C)
                  \pmod{2\pi}.
 \label{eq:spinone-info-holonomy}
\end{equation}

\subsubsection*{Uncertainty relation and quantum geometric tensor}

We briefly review the quantum geometric tensor, which unifies the information metric and the Berry curvature. Let us define
\begin{equation}
    |D_i\psi\rangle=|\partial_i\psi\rangle-\langle\psi|\partial_i\psi\rangle|\psi\rangle.
\end{equation}
The quantum geometric tensor is defined as
\begin{equation}
    Q_{ij}:=  \langle D_i\psi|D_j\psi\rangle=\langle \partial_i\psi|\partial_j\psi\rangle-\langle\psi|\partial_j\psi\rangle\langle\partial_i\psi|\psi\rangle.
\end{equation}
We can check that it satisfies the following properties:
\begin{equation}
    Q_{ij}=Q_{ji}^*,\quad Q_{ij}=2g_{ij}^B-\frac{i}{2}\mathcal{F}_{ij}.
\end{equation}
Since $Q_{ij}$ is a matrix of inner products of states, it is positive semidefinite:
\begin{equation}
    0\leq \operatorname{det}Q_{ij}=4\operatorname{det}g_{ij}^B-\frac{\mathcal{F}_{12}^2}{4}.
\end{equation}
Thus, $|\mathcal{F}_{12}|\leq 4\sqrt{\operatorname{det}g_{ij}^B}$.
For $|\psi_q\rangle$, $g^B_{ij}=g^{B,q}_{ij}$ and $\mathcal{F}_q=\frac{1}{2}\mathcal{F}_{ij}\,dx^i\wedge dx^j$, with $(x^1,x^2)=(\rho,\phi)$.
In particular, in our case, we find
\begin{equation}
    \operatorname{det}Q_{ij}=\left(V_q^2-\frac{q^2}{4}\right)\sinh^2\rho\geq0,
\end{equation}
which gives $V_q\geq \frac{|q|}{2}$.

\subsubsection{Coefficient of the information metric}
\label{sec:spinone-info-coefficient}

We now evaluate the Bures coefficient $\mathcal G_q=V_q/2$ from the descendant expansion in eqs.\,\eqref{eq:ads-state-plus}--\eqref{eq:ads-state-minus}. Let $d=1-q$ be the level difference in $|k,k+d\rangle_+$. We now evaluate the overlap of the regulated bulk local states. Their squared norms are
\begin{equation}
    ||e^{-\varepsilon H}|\Phi_q^{(+)}\rangle||^2=|\mathcal{N}_A|^2e^{-2\varepsilon(\Delta+d)}\sum_{k=0}^\infty f_q(k)e^{-4\varepsilon k}.
\end{equation}
To simplify the notation, we define
\begin{equation}
    z=e^{-4\varepsilon},\quad Z_q(z)=\sum_{k=0}^\infty f_q(k)z^k.\label{z}
\end{equation}
The explicit form of $f_q(k)$ is given by
\begin{equation}
 f_1(k)=\frac{(\Delta+k)(\Delta+k-1)}{\Delta(\Delta-1)},\quad
 f_0(k)=\frac{2(k+1)(\Delta+k)}{\Delta(\Delta-1)},\quad
 f_{-1}(k)=\frac{(k+1)(k+2)}{\Delta(\Delta-1)}.
 \label{eq:spinone-info-f}
\end{equation}
By direct computation, we can write $Z_q(z)$ as
                \begin{align}
                    Z_1(z)&=\frac{\Delta(\Delta-1)+2\Delta(2-\Delta)z+(\Delta-1)(\Delta-2)z^2}{\Delta(\Delta-1)(1-z)^3},
                    \nonumber\\
                    Z_0(z)&=\frac{2[\Delta+(2-\Delta)z]}{\Delta(\Delta-1)(1-z)^3},
                    \qquad
                    Z_{-1}(z)=\frac{2}{\Delta(\Delta-1)(1-z)^3}.
                    \label{eq:spinone-info-Z}
                \end{align}

Let us compute $V_q=\langle K_1^2\rangle=\langle K_2^2\rangle$. For technical convenience, we consider
\begin{equation}
    V_q=\frac{1}{2}\langle K_1^2+K_2^2\rangle.
\end{equation}
The normalized vector $|v_q\rangle$ can also be written as
\begin{equation}
    |v_q\rangle=\sum_{k=0}^\infty u_k|k,k+d\rangle_+,\quad u_k=\frac{(-1)^k\sqrt{f_q(k)}z^{\frac{k}{2}}}{\sqrt{Z_q(z)}}.
\end{equation}
Using the commutation relations, we obtain
\begin{align}
 K_1^2+K_2^2={}&\mathcal{L}_0^2-h(h-1)
       +\overline{\mathcal{L}}_0^2-\bar h(\bar h-1)
 \nonumber\\
 &-\bigl(\mathcal{L}_1\overline{\mathcal{L}}_1
       +\mathcal{L}_{-1}\overline{\mathcal{L}}_{-1}\bigr),
 \label{eq:spinone-info-Ksum}
\end{align}
where $2h=\Delta+1$ and $2\bar{h}=\Delta-1$.
We define the coefficients as
\begin{equation}
    (K_1^2+K_2^2)|k,k+d\rangle_+=D_k|k,k+d\rangle_+-r_k|k-1,k+d-1\rangle_+-r_{k+1}|k+1,k+d+1\rangle_+.
\end{equation}
Let us first compute $D_k$. Using
\begin{align}
    \mathcal{L}_0|k,k+d\rangle_+&=(h+k)|k,k+d\rangle_+,\nonumber\\
    \overline{\mathcal{L}}_0|k,k+d\rangle_+&=(\bar h+k+d)|k,k+d\rangle_+,
\end{align}
we find
\begin{equation}
    D_k=2k^2+2(\Delta+d)k+\Delta(d+1)+d(d-1).
\end{equation}
Similarly, using
\begin{align}
     \mathcal{L}_1|k,k+d-1\rangle_+&=\sqrt{k(2h+k-1)}|k-1,k+d-1\rangle_+,\nonumber\\
    \overline{\mathcal{L}}_1|k,k+d\rangle_+&=\sqrt{(k+d)(2\bar h+k+d-1)}|k,k+d-1\rangle_+,
\end{align}
we obtain
\begin{equation}
    r_k=\sqrt{k(\Delta+k)(k+d)(\Delta+k+d-2)},\quad r_0=0.
\end{equation}
We normalize the states as ${}_+\langle j,j+d|k,k+d\rangle_+=\delta_{jk}$. With this convention, $V_q$ is given by
\begin{align}
    2V_q&=\sum_{k=0}^\infty D_k|u_k|^2-\sum_{k=1}^\infty r_ku_{k-1}^* u_k-\sum_{k=0}^\infty r_{k+1}u_{k+1}^*u_k\nonumber\\
    &=\sum_{k=0}^\infty D_k|u_k|^2-2\Re \left(\sum_{k=1}^\infty r_ku_{k-1}^* u_k\right).
\end{align}
Using the explicit form of $u_k$, we can further simplify the result as
\begin{equation}
    V_q=\frac{1}{Z_q}\sum_{k=0}^{\infty}f_q(k)z^k\left((1+e^{2\varepsilon})k^2+(\Delta+d+e^{2\varepsilon}(\Delta+d-2))k+\frac{\Delta(d+1)+d(d-1)}{2}\right).
\end{equation}
Using eq.\,\eqref{z}, we obtain
\begin{equation}
    \sum_k kf_q(k)z^k=zZ_q'(z),\quad \sum_k k^2f_q(k)z^k=zZ_q'(z)+z^2Z_q''(z),
\end{equation}
where primes denote derivatives with respect to $z$. Substituting these expressions, we obtain
\begin{equation}
    V_q=(1+e^{2\varepsilon})\frac{zZ_q'+z^2Z_q''}{Z_q}+(\Delta+d+e^{2\varepsilon}(\Delta+d-2))\frac{zZ_q'}{Z_q}+\frac{\Delta(d+1)+d(d-1)}{2},
\end{equation}
where $Z_q$ is given in eq.\,\eqref{eq:spinone-info-Z}.
For example,
                \begin{equation}
                    V_{-1}=\frac{3\Delta+2+2e^{-2\varepsilon}+(8-3\Delta)e^{-4\varepsilon}}{2(1-e^{-2\varepsilon})^2}.
                    \label{eq:spinone-info-Vminus}
                \end{equation}
Thus, the coefficients are not universally independent of polarization. At fixed $\Delta$, the large-$k$ behavior $f_q(k)\propto k^2$ gives
\begin{equation}
 \mathcal G_q=\frac{3}{4\varepsilon^2}+O(\varepsilon^{-1}),
 \qquad \varepsilon\to0,
 \label{eq:spinone-info-UV}
\end{equation}
with the same leading divergence for every component. The negative-helicity branch follows by interchanging the two modules: its metric coefficient at fixed $q$ equals $\mathcal G_{-q}$ above. Its Berry curvature remains eq.\,\eqref{eq:spinone-info-berry} because the rotation eigenvalue is still $q$.

\subsubsection{Flat limit at fixed physical resolution}
\label{sec:spinone-info-flat}
Let us consider the flat limit of the metric.
A finite physical smearing time requires $\tau_c=\ell\varepsilon$ to be fixed. We therefore take
\begin{equation}
 \Delta=1+m\ell,\qquad \varepsilon=\frac{\tau_c}{\ell},
 \qquad r=\ell\rho,
 \qquad \ell\to\infty\quad(m,\tau_c,r\ \text{fixed}),
 \label{eq:spinone-info-flat-limit}
\end{equation}
with $m,\tau_c>0$, where $r$ is the physical radial distance.
Keeping $\varepsilon$ fixed instead would send the physical smearing time to infinity and describe a different family of states.

The relevant levels grow as $k=O(\ell)$. To make the limit explicit, define the physical energy
$E_k=(\Delta+2k+d)/\ell$, whose spacing is $2/\ell$. At fixed $E>m$, eq.\,\eqref{eq:spinone-info-f} has the continuum limit
\begin{equation}
 f_q(k)\longrightarrow\frac{w_q(E)}{4m^2},\qquad
 \begin{cases}
 w_1(E)=(E+m)^2,\\
 w_0(E)=2(E^2-m^2),\\
 w_{-1}(E)=(E-m)^2.
 \end{cases}
 \label{eq:spinone-info-weights}
\end{equation}
Thus the regulated sum, without $|\mathcal{N}_A|^2$, behaves as
\begin{equation}
 \sum_{k=0}^{\infty}f_q(k)e^{-2\tau_c E_k}
 \sim\frac{\ell}{8m^2}
      \int_m^\infty w_q(E)e^{-2\tau_c E}\,dE.
 \label{eq:spinone-info-riemann}
\end{equation}
For fixed positive $m$ and $\tau_c$, the sums converge as Riemann sums on any finite energy interval. For sufficiently large $\ell$, the remaining tails, including the energy moments used below, are uniformly bounded by a polynomial in $E$ times $e^{-2\tau_c E}$. We can therefore take the limit of the normalized distribution and its first two energy moments. The expectation value in the flat limit is given by
\begin{equation}
    \langle\mathcal{O}\rangle=\frac{\int_m^\infty\mathcal{O}w_q(E) e^{-2\tau_cE}dE}{\int_m^\infty w_q(E) e^{-2\tau_cE}dE}.
\end{equation}
In the flat limit, $V_q$ reads
\begin{equation}
    V_q=2\langle k ^2+m\ell k\rangle+\mathcal{O}(\ell),
\end{equation}
which leads to
\begin{equation}
 \mathcal{G}_q^{\mathrm{flat}}:=\lim_{\ell\to\infty}\frac{\mathcal G_q}{\ell^2}
     =\frac14\langle E^2-m^2\rangle_q
     =\frac14\langle\boldsymbol{p}^{2}\rangle_q.
 \label{eq:spinone-info-flat-variance}
\end{equation}
Rotational invariance gives
$\langle p_x^2\rangle=\langle p_y^2\rangle=\langle\boldsymbol p^2\rangle/2$.

For completeness, let us evaluate $\mathcal{G}_q^{\mathrm{flat}}$ explicitly. We use $x=2m\tau_c$ and $y=E-m$. Rewriting the integral, we obtain
\begin{equation}
    \mathcal{G}_q^{\mathrm{flat}}=\frac{J_q}{4I_q},
\end{equation}
where
\begin{equation}
    I_q:=\int_0^\infty w_q(m+y)e^{-2\tau_c y}dy,\quad J_q:=\int_0^\infty (y^2+2my)w_q(m+y)e^{-2\tau_c y}dy.
\end{equation}
Evaluating the integrals, we obtain
\begin{equation}
 I_1=\frac{2}{(2\tau_c)^3}(1+2x+2x^2),\qquad
 I_0=\frac{4}{(2\tau_c)^3}(1+x),\qquad I_{-1}=\frac{2}{(2\tau_c)^3},
 \label{eq:spinone-info-flat-norms}
\end{equation}
and
\begin{align}
 J_1&=\frac{24}{(2\tau_c)^5}+\frac{36m}{(2\tau_c)^4}
                    +\frac{24m^2}{(2\tau_c)^3}+\frac{8m^3}{(2\tau_c)^2},
 \nonumber\\
 J_0&=\frac{48}{(2\tau_c)^5}+\frac{48m}{(2\tau_c)^4}+\frac{16m^2}{(2\tau_c)^3},
 \qquad J_{-1}=\frac{24}{(2\tau_c)^5}+\frac{12m}{(2\tau_c)^4}.
 \label{eq:spinone-info-flat-numerators}
\end{align}
The definition $\mathcal{G}_q^{\mathrm{flat}}=J_q/(4I_q)$ gives
\begin{align}
 \mathcal{G}_1^{\mathrm{flat}}&=\frac{6+9x+6x^2+2x^3}{2(2\tau_c)^2(1+2x+2x^2)},
 \nonumber\\
 \mathcal{G}_0^{\mathrm{flat}}&=\frac{6+6x+2x^2}{2(2\tau_c)^2(1+x)},\qquad
 \mathcal{G}_{-1}^{\mathrm{flat}}=\frac{6+3x}{2(2\tau_c)^2}.
 \label{eq:spinone-info-flat-V}
\end{align}
Thus, in the flat limit, the geometry itself contracts to
\begin{equation}
 ds_{\mathrm B,q}^2
   \longrightarrow \mathcal{G}_q^{\mathrm{flat}}(dr^2+r^2d\phi^2),
 \qquad
 \mathcal{F}_q=\left(\frac{qr}{\ell^2}+O(\ell^{-4})\right)
           dr\wedge d\phi\longrightarrow0.
 \label{eq:spinone-info-flat-geometry}
\end{equation}

Although the position-space Berry curvature disappears, the metric retains component-dependent information at finite resolution. For $m\tau_c\ll1$, every component has $\mathcal{G}_q^{\mathrm{flat}}\sim3/(4\tau_c^2)$. In the nonrelativistic regime $m\tau_c\gg1$, however,
\begin{equation}
 \mathcal{G}_1^{\mathrm{flat}}\sim\frac{m}{4\tau_c},\qquad
 \mathcal{G}_0^{\mathrm{flat}}\sim\frac{m}{2\tau_c},\qquad
 \mathcal{G}_{-1}^{\mathrm{flat}}\sim\frac{3m}{4\tau_c}.
 \label{eq:spinone-info-nonrelativistic}
\end{equation}
The ratio $1:2:3$ follows directly from the threshold behavior in eq.\,\eqref{eq:spinone-info-weights}. With $E-m\simeq\boldsymbol{p}^{2}/(2m)$, the three weights are proportional to $p^0,p^2,p^4$, respectively, while the regulator becomes $e^{-\tau_c p^2/m}$. The two-dimensional radial measure is proportional to $p\,dp$ at this order. Weighting it by $p^{2n}$ gives $\langle p^2\rangle=(n+1)m/\tau_c$, with $n=0,1,2$, reproducing eq.\,\eqref{eq:spinone-info-nonrelativistic}.

    \section{Carrollian holography}
        \label{sec:intrinsic}
We then consider the similar procedure to reconstruct the local spin-1 state in the Carrollian holography in flat spacetime.
        The construction in this section uses only the Carrollian symmetry algebra and is largely independent of the AdS analysis in Section~\ref{sec:ads}, and the comparison between the two via the flat limit is deferred to Section~\ref{sec:flat-limit}.

        \subsection{Overview of BMS symmetry and Carrollian holography}
            Carrollian holography seeks to describe gravity in asymptotically flat spacetime by a quantum field theory on null infinity \cite{Bagchi:2010zz,Bagchi:2012cy,Barnich:2010eb}.
            Unlike the timelike boundary of AdS, null infinity has a degenerate intrinsic metric, so the proposed boundary theory has Carrollian rather than relativistic spacetime symmetry.
            For the three-dimensional Minkowski spacetime considered below, the metric in retarded coordinates is \cite{Hao:2025btl}
            \begin{align}
                ds^2=-du^2-2du\,dr+r^2d\phi^2,
                \qquad u=t-r,\qquad \phi\sim\phi+2\pi.
                \label{eq:carrollian-retarded-metric}
            \end{align}
            Here $r\geq0$, and the Cartesian spatial coordinates are $x=r\cos\phi$ and $y=r\sin\phi$.
            Future null infinity is reached as $r\to\infty$ at fixed $(u,\phi)$ and has the topology of a cylinder.
            The pullback of the conformally rescaled metric $r^{-2}ds^2$ to this boundary is $d\phi^2$, whose kernel is spanned by $\partial_u$ \cite{Duval:2014carroll}.
            The boundary theory retains this time coordinate, whereas celestial holography organizes scattering amplitudes as correlators on the celestial sphere \cite{Pasterski:2016qvg,Strominger:2017zoo}.
            Tests of the Carrollian proposal include cosmological horizon entropy \cite{Barnich:2012xq, Bagchi:2012xr}, entanglement entropy \cite{Bagchi:2014iea, Jiang:2017ecm, Apolo:2020bld, Hao:2025naz}, and scattering correlators \cite{Salzer:2023jqv}.

            The asymptotic symmetries of the bulk form the Bondi--Metzner--Sachs (BMS) group \cite{Bondi:1962px,Sachs:1962wk,Barnich:2006av}.
            In three dimensions it acts on the boundary coordinates as $\phi\mapsto f(\phi)$ and $u\mapsto f'(\phi)u+g(\phi)$, where $f$ is a reparametrization of the circle and $g$ is a periodic function.
            The generators $L_n$ describe the angular reparametrizations, called superrotations, while $M_n$ generate the angle-dependent time shifts called supertranslations.
            Their charges obey the centrally extended BMS$_3$ algebra
            \begin{align}
                \begin{aligned}
                    [L_m,L_n]&=(m-n)L_{m+n}+\frac{c_L}{12}m(m^2-1)\delta_{m+n,0},\\
                    [L_m,M_n]&=(m-n)M_{m+n}+\frac{c_M}{12}m(m^2-1)\delta_{m+n,0},\\
                    [M_m,M_n]&=0,
                \end{aligned}
                \qquad m,n\in\mathbb{Z}.
                \label{eq:bms3-review-algebra}
            \end{align}
            Here $c_L$ and $c_M$ are central charges, with $c_L=0$ and $c_M=3/G$ for three-dimensional Einstein gravity, where $G$ is Newton's constant \cite{Barnich:2006av}.
            The six generators $L_{0,\pm1}$ and $M_{0,\pm1}$ form the global subalgebra $\mathfrak{iso}(2,1)$, the Poincar\'{e} algebra of the bulk.
            The central terms vanish in this subalgebra, which is sufficient for the free-particle reconstruction below.
            On the Minkowski background of eq.\,\eqref{eq:carrollian-retarded-metric}, these generators are realized by the six Killing vector fields \cite{Hao:2025btl}
            \begin{align}
                L_0&=i\partial_\phi,\qquad M_0=i\partial_u,
                \nonumber\\
                L_{\pm1}&=e^{\pm i\phi}\left(-i u\partial_u+i(u+r)\partial_r
                \mp\left(1+\frac{u}{r}\right)\partial_\phi\right),
                \nonumber\\
                M_{\pm1}&=e^{\pm i\phi}\left(\mp\partial_u\pm\partial_r+\frac{i}{r}\partial_\phi\right).
                \label{eq:carrollian-bulk-killing}
            \end{align}
            In Cartesian coordinates, $L_0$ generates spatial rotations and $L_{\pm1}$ are linear combinations of boosts, while $M_{0,\pm1}$ generate translations.

            To construct boundary states, one must also choose a representation of this algebra.
            We label a primary by $L_0|\Delta,\xi\rangle=\Delta|\Delta,\xi\rangle$ and $M_0|\Delta,\xi\rangle=\xi|\Delta,\xi\rangle$.
            A highest weight primary is annihilated by the positive modes, as in the BMS bootstrap \cite{Bagchi:2017cpu}, whereas an induced representation is built from supertranslation eigenstates \cite{Barnich:2014kra,Oblak:2015sea,Campoleoni:2016vsh}.
            For the global algebra, the respective conditions are
            \begin{align}
                L_1|\Delta,\xi\rangle_{\mathrm{HW}}
                &=M_1|\Delta,\xi\rangle_{\mathrm{HW}}=0,
                \nonumber\\
                M_{1}|\Delta,\xi\rangle
                &=M_{-1}|\Delta,\xi\rangle=0.
                \label{eq:bms3-review-representations}
            \end{align}
            The second line describes a rest-frame state with energy $\xi$ and rotation weight $\Delta$, from which Lorentz transformations generate moving states.
            These induced representations arise in the flat limit of unitary AdS representations.
            Reconstruction using highest weight primaries has been studied for scalar and vector fields \cite{Chen:2023naw}.
            For massive scalar bulk local states, the induced representation admits solutions whose overlaps reproduce the bulk propagator, whereas the highest weight ansatz encounters an obstruction \cite{Hao:2025btl}.
            The scalar construction also has an explicit flat limit from AdS \cite{Hao:2025btl}.
            A subsequent analysis supplies the dual basis needed to take that limit consistently and extends the reconstruction to higher dimensions \cite{Hao:2026cqm}.
            We use the induced representation to construct the vector states below and examine the highest weight obstruction separately.

\subsection{Constraint equations}
        \label{subsec:carrollian-constraints}

            Let $\ket{\Phi_q(t)}:=\ket{\Phi_q(t,0,0)}$ denote the helicity-$q$ component of a massive
            spin-one bulk local state at the bulk point $(t,0,0)$, with $q=+1,0,-1$.  We use the normalized
            spin-one basis
            \begin{align}
                S_0\ket{q}&=q\ket{q},\label{eq:spin1-S0}\\
                S_+\ket{-1}&=\sqrt{2}\ket{0},\qquad
                S_+\ket{0}=\sqrt{2}\ket{+1},\qquad
                S_+\ket{+1}=0,\label{eq:spin1-Splus}\\
                S_-\ket{+1}&=\sqrt{2}\ket{0},\qquad
                S_-\ket{0}=\sqrt{2}\ket{-1},\qquad
                S_-\ket{-1}=0.\label{eq:spin1-Sminus}
            \end{align}

            The stabilizer generators at $(t,0,0)$ are
            \begin{align}
                A_t&:=L_1-i t M_1,
                &
                B_t&:=L_{-1}+i t M_{-1}.
                \label{eq:At-Bt}
            \end{align}
            Using the global BMS$_3$ algebra, one finds
            \begin{align}
                [L_0,A_t]&=-A_t,
                &
                [L_0,B_t]&=B_t,
                &
                [A_t,B_t]&=2L_0.
                \label{eq:AtBt-algebra}
            \end{align}
            The local covariance conditions are
            \begin{align}
                L_0\ket{\Phi_q(t)}&=q\ket{\Phi_q(t)},
                \label{eq:intrinsic-L0}\\
                A_t\ket{\Phi_q(t)}&=S_-\ket{\Phi_q(t)},
                \label{eq:intrinsic-A}\\
                B_t\ket{\Phi_q(t)}&=-S_+\ket{\Phi_q(t)}.
                \label{eq:intrinsic-B}
            \end{align}
            Equivalently,
            \begin{align}
                A_t\ket{\Phi_{+1}}&=\sqrt{2}\ket{\Phi_0},
                &
                A_t\ket{\Phi_0}&=\sqrt{2}\ket{\Phi_{-1}},
                &
                A_t\ket{\Phi_{-1}}&=0,
                \label{eq:component-A}\\
                B_t\ket{\Phi_{+1}}&=0,
                &
                B_t\ket{\Phi_0}&=-\sqrt{2}\ket{\Phi_{+1}},
                &
                B_t\ket{\Phi_{-1}}&=-\sqrt{2}\ket{\Phi_0}.
                \label{eq:component-B}
            \end{align}
            The commutator $[A_t,B_t]=2L_0$ requires opposite signs in the two mixing equations.
            Their assignment here agrees with the AdS component convention in eqs.\,\eqref{eq:ads-local-L1} and \eqref{eq:ads-local-L-1}.

        \subsection{Solution in the induced representation}
        \label{subsec:carrollian-induced-solution}

            We now solve the constraints in the induced representation.  An induced primary satisfies
            \begin{align}
                L_0\ket{\Delta,\xi}&=\Delta\ket{\Delta,\xi},
                &
                M_0\ket{\Delta,\xi}&=\xi\ket{\Delta,\xi},
                &
                M_{\pm1}\ket{\Delta,\xi}&=0,
                \label{eq:induced_rep}
            \end{align}
            and for a massive excitation we set
            \begin{align}
                \xi=m.
                \label{eq:xi-mass}
            \end{align}
            The two massive spin-one branches have labels
            \begin{align}
                (\Delta,\xi)=(+1,m),
                \qquad
                (\Delta,\xi)=(-1,m).
                \label{eq:two-flat-branches}
            \end{align}
            The two branches are distinct induced-module solutions.  They will be related by parity below, but
            neither branch is introduced only as a parity completion of the other.

            The order of the generators in the descendant basis is important.  We therefore distinguish the two
            ordered bases
            \begin{align}
                \ket{i,j}_{-+}^{\Delta,\xi}
                &:=L_{-1}^{\,i}L_1^{\,j}\ket{\Delta,\xi},
                \label{eq:basis-minus-plus}\\
                \ket{i,j}_{+-}^{\Delta,\xi}
                &:=L_1^{\,i}L_{-1}^{\,j}\ket{\Delta,\xi},
                \label{eq:basis-plus-minus}
            \end{align}
            with $i,j\in\mathbb Z_{\geq0}$.  Since $[L_1,L_{-1}]=2L_0$, the two bases are not related by a trivial relabeling for $\Delta \ne 0$.
            For scalar states with $\Delta=0$, these two bases are equivalent.

            In the first basis the useful actions are
            \begin{align}
                L_{-1}\ket{i,j}_{-+}
                &=\ket{i+1,j}_{-+},\nonumber\\
                L_0\ket{i,j}_{-+}
                &=(\Delta+i-j)\ket{i,j}_{-+},\nonumber\\
                L_1\ket{i,j}_{-+}
                &=\ket{i,j+1}_{-+}
                +\bigl[i(i-1)+2i(\Delta-j)\bigr]\ket{i-1,j}_{-+},\nonumber\\
                M_{-1}\ket{i,j}_{-+}
                &=-2j\xi\ket{i,j-1}_{-+},\nonumber\\
                M_1\ket{i,j}_{-+}
                &=2i\xi\ket{i-1,j}_{-+}
                -2i(i-1)j\xi\ket{i-2,j-1}_{-+}.
                \label{eq:induced-actions-minus-plus}
            \end{align}
            States with negative descendant labels are understood to vanish.  The corresponding formulas in the
            $L_1^iL_{-1}^j$ basis follow from the same algebra and are the natural ones for the opposite branch.

            \paragraph{Branch I: $(\Delta,\xi)=(+1,m)$.}
            In the $L_{-1}^iL_1^j$ basis, the $L_0$ condition fixes the support to $j=i+r$ for the component
            $q=1-r$, with $r=0,1,2$.
            With $\alpha(t)=-i/(2mt)$, substitution into the local covariance conditions gives
            \begin{align}
                \ket*{\Phi^{(+)}_{1-r}(t,0,0)}
                =\mathcal N_+(t)
                \sqrt{\binom{2}{r}}\,
                \big(\alpha(t)\big)^r
                \sum_{k=0}^{\infty}\frac{\alpha(t)^k}{k!}
                \ket{k,k+r}_{-+}^{1,m},
                \qquad r=0,1,2.
                \label{eq:flat-compact}
            \end{align}
            Equivalently,
            \begin{align}
                \ket*{\Phi^{(+)}_{+1}(t)}
                &=\mathcal N_+(t)
                \sum_{k=0}^{\infty}\frac{1}{k!}
                \left(-\frac{i}{2mt}\right)^k
                \ket*{k,k}_{-+}^{1,m},
                \label{eq:flat-plus}\\
                \ket*{\Phi^{(+)}_0(t)}
                &=-\mathcal{N}_+(t)\frac{i}{\sqrt{2}\,mt}
                \sum_{k=0}^{\infty}\frac{1}{k!}
                \left(-\frac{i}{2mt}\right)^k
                \ket*{k,k+1}_{-+}^{1,m},
                \label{eq:flat-zero}\\
                \ket*{\Phi^{(+)}_{-1}(t)}
                &=-\mathcal N_+(t)\frac{1}{4m^2t^2}
                \sum_{k=0}^{\infty}\frac{1}{k!}
                \left(-\frac{i}{2mt}\right)^k
                \ket*{k,k+2}_{-+}^{1,m}.
                \label{eq:flat-minus}
            \end{align}
            The stabilizer equations determine all relative coefficients but leave the common function
            $\mathcal N_+(t)$ undetermined.  The time-evolution equation
            \begin{align}
                \partial_t\ket{\Phi_q(t)}
                =i M_0\ket{\Phi_q(t)}
                \label{eq:local-time-evolution}
            \end{align}
            fixes its time dependence to
            \begin{align}
                \mathcal N_+(t)=A_+\frac{e^{i mt}}{t},
                \label{eq:plus-normalization-function}
            \end{align}
            with $A_+$ independent of $t$.

            \paragraph{Branch II: $(\Delta,\xi)=(-1,m)$.}
            In this module it is more natural to use the opposite ordered basis $L_1^iL_{-1}^j$.  The
            $L_0$ condition then fixes $j=i+r$ for the component $q=-1+r$.
            Solving the covariance equations directly gives the symmetric counterpart
            \begin{align}
                \ket*{\Phi^{(-)}_{-1+r}(t,0,0)}
                =\mathcal N_-(t)
                \sqrt{\binom{2}{r}}\,
                \big(-\alpha(t)\big)^r
                \sum_{k=0}^{\infty}\frac{\alpha(t)^k}{k!}
                \ket{k,k+r}_{+-}^{-1,m},
                \qquad r=0,1,2.
                \label{eq:flat-compact-minus}
            \end{align}
            Explicitly,
            \begin{align}
                \ket*{\Phi^{(-)}_{-1}(t)}
                &=\mathcal N_-(t)
                \sum_{k=0}^{\infty}\frac{1}{k!}
                \left(-\frac{i}{2mt}\right)^k
                \ket*{k,k}_{+-}^{-1,m},
                \label{eq:flat-minus-branch-minus}\\
                \ket*{\Phi^{(-)}_0(t)}
                &=\mathcal{N}_-(t)\frac{i}{\sqrt{2}\,mt}
                \sum_{k=0}^{\infty}\frac{1}{k!}
                \left(-\frac{i}{2mt}\right)^k
                \ket*{k,k+1}_{+-}^{-1,m},
                \label{eq:flat-minus-branch-zero}\\
                \ket*{\Phi^{(-)}_{+1}(t)}
                &=-\mathcal N_-(t)\frac{1}{4m^2t^2}
                \sum_{k=0}^{\infty}\frac{1}{k!}
                \left(-\frac{i}{2mt}\right)^k
                \ket*{k,k+2}_{+-}^{-1,m}.
                \label{eq:flat-minus-branch-plus}
            \end{align}
            The common function is fixed by the time-evolution equation as
            \begin{align}
                \mathcal{N}_-(t)=A_-\frac{e^{i mt}}{t}.
                \label{eq:minus-normalization-function}
            \end{align}

            The two formulas are related by the parity automorphism
            \begin{align}
                \mathsf P(L_n)&=-L_{-n},
                &
                \mathsf P(M_n)&=M_{-n},
                \label{eq:parity-algebra}
            \end{align}
            which maps
            \begin{align}
                \mathsf P:\quad
                \Delta&\longmapsto-\Delta,
                &
                \xi&\longmapsto\xi,
                &
                A_t&\longmapsto-B_t,
                &
                B_t&\longmapsto-A_t.
                \label{eq:parity-map}
            \end{align}
            On the ordered descendant bases,
            \begin{align}
                \mathsf P\ket{i,j}_{-+}^{\Delta,\xi}
                =(-1)^{i+j}\ket{i,j}_{+-}^{-\Delta,\xi}.
                \label{eq:parity-basis-map}
            \end{align}
            Hence the two independently solved branches obey the exact relation
            \begin{align}
                \ket*{\Phi_q^{(-)}(t)}
                =\mathsf P\ket*{\Phi_{-q}^{(+)}(t)}.
                \label{eq:parity-state-relation}
            \end{align}
            This relation is a consistency check between the two branch solutions, not a replacement for the
            second solution.

            For the inner-product calculation below we Wick rotate to Euclidean time,
            \begin{align}
                t\longrightarrow i\tau,
                \qquad
                \tau>0,
                \qquad
                s:=-2m\tau.
                \label{eq:wick-rotation}
            \end{align}
            For the $\Delta=-1$ branch, changing the descendant ordering introduces an additional factor $\alpha(i\tau)^2$ in the common prefactor.
            We write this reordered prefactor as
            \begin{align}
                \mathcal{N}_-(i\tau)\alpha(i\tau)^2
                =\mathcal{B}_0\frac{e^{-m\tau}}{\tau^3},
                \label{eq:B0-normalization-function}
            \end{align}
            where $\mathcal{B}_0$ is fixed below by matching the scalar Green-function normalization.

            Although the natural basis of this branch is $L_1^iL_{-1}^j$, the dual bra used below is simplest in
            the $L_{-1}^iL_1^j$ basis.  Re-expanding the same $\Delta=-1$ states in that ordered basis gives
            \begin{align}
                \ket*{\Phi_q^{(-)}(\tau,0,0)}
                =
                \mathcal B_0\frac{e^{-m\tau}}{\tau^3}
                \sum_{k=0}^{\infty}
                \frac{\mathcal C_q(k,s)}{s^k k!}
                \ket{k+q+1,k}_{-+}^{-1,m},
                \label{eq:negative-reordered-states}
            \end{align}
            with
            \begin{align}
                \mathcal C_{+1}(k,s)&=1,\nonumber\\
                \mathcal C_0(k,s)&=\sqrt2\,(k+2-s),\nonumber\\
                \mathcal C_{-1}(k,s)&=
                s^2-2(k+1)s+(k+1)(k+2).
                \label{eq:C-coefficients}
            \end{align}
            These polynomials are not a second ansatz for the state.  They are generated solely by reordering the
            noncommuting $L_1$ and $L_{-1}$.  For example,
            \begin{align}
                L_1L_{-1}^{\,n}
                =
                L_{-1}^{\,n}L_1
                +nL_{-1}^{\,n-1}(2L_0+n-1),
                \label{eq:generator-reordering}
            \end{align}
            and repeated use of this identity produces
            \eqref{eq:C-coefficients}.

        \subsection{Dual basis}
        \label{subsec:carrollian-dual-basis}

            For the pairing we use the algebraic dual of the ordered basis
            $\ket{i,j}_{-+}^{-1,m}=L_{-1}^iL_1^j\ket{-1,m}$.  Its defining pairing is
            \begin{align}
                \bra*{\widetilde{i,j}}k,l\rangle_{-+}
                =
                \delta_{ik}\delta_{jl}.
                \label{eq:dual-basis-pairing}
            \end{align}
            This is an algebraic dual basis rather than the Hermitian conjugate of the ket basis.

            For $\Delta=-1$, the three local bras at the origin may be chosen as
            \begin{align}
                \bra{\Phi_{-1}(0)}&=\bra*{\widetilde{0,0}},\label{eq:bra-minus}\\
                \bra{\Phi_0(0)}&=\sqrt{2}\,\bra*{\widetilde{1,0}},\label{eq:bra-zero}\\
                \bra{\Phi_{+1}(0)}&=-\bra*{\widetilde{1,0}}L_1.\label{eq:bra-plus}
            \end{align}
            Indeed, using the $\Delta=-1$ action of $L_1$,
            \begin{align}
                -\bra*{\widetilde{1,0}}L_1\ket{i,j}_{-+}^{-1,m}
                =
                2\delta_{i2}\delta_{j0}.
                \label{eq:bra-plus-action}
            \end{align}
            Thus these three bras extract precisely the three coefficient slots selected by the local spin-one
            multiplet.  Their right actions obey
            \begin{align}
                \bra{\Phi_q}L_0&=q\bra{\Phi_q},\nonumber\\
                \bra{\Phi_{-1}}L_1&=-\sqrt2\bra{\Phi_0},
                &
                \bra{\Phi_0}L_1&=-\sqrt2\bra{\Phi_{+1}},
                &
                \bra{\Phi_{+1}}L_1&=0,\nonumber\\
                \bra{\Phi_{-1}}L_{-1}&=0,
                &
                \bra{\Phi_0}L_{-1}&=\sqrt2\bra{\Phi_{-1}},
                &
                \bra{\Phi_{+1}}L_{-1}&=\sqrt2\bra{\Phi_0}.
                \label{eq:dual-right-actions}
            \end{align}

            The $\Delta=+1$ branch is naturally treated in the opposite ordered basis
            $L_1^iL_{-1}^j$.  In that basis the analogous finite dual construction is the parity counterpart of
            \eqref{eq:bra-minus}--\eqref{eq:bra-plus}.  We therefore do not duplicate the calculation here:
            the $\Delta=+1$ local state and its dual are already distinct branch solutions, while parity relates
            their pairings exactly.  For the explicit Green-function calculation it is sufficient to work with the
            $\Delta=-1$ branch in the ordering in which the dual basis is finite and simple.

            We define the local-state inner-product matrix in the helicity ordering $(+1,0,-1)$ by
            \begin{align}
                I^{(-)}_{qq'}(X)
                &:={}_{-1,m}\!\bra{\Phi_q(0)}
                \Phi_{q'}(X)\rangle_{-1,m},
                \qquad q,q'=+1,0,-1.
                \label{eq:I-definition}
            \end{align}

        \subsection{Bulk propagator from inner product of the states}
        \label{subsec:carrollian-bulk-propagator}

            We evaluate the inner-product matrix in eq.\,\eqref{eq:I-definition} in Euclidean signature, using the coordinates
            \begin{align}
                X&=(\tau,x,y),
                &
                X'&=(0,0,0),\nonumber\\
                \rho^2&:=x^2+y^2,
                &
                R&:=\sqrt{\tau^2+\rho^2},
                &
                z&:=x-i y,
                \qquad
                \bar z:=x+i y.
                \label{eq:euclidean-coordinates}
            \end{align}
            We translate the state in the spatial directions by acting with the supertranslation generators,
            \begin{align}
                \ket{\Phi_q(\tau,x,y)}
                =
                e^{(zM_1-\bar{z}M_{-1})/2}
                \ket{\Phi_q(\tau,0,0)}.
                \label{eq:spatial-translation}
            \end{align}

            We calculate the translation matrix elements using the generating vector in the $L_{-1}^iL_1^j$ basis,
            \begin{align}
                \mathcal F(u,v)
                :=
                \sum_{i,j\geq0}
                \ket{i,j}_{-+}^{-1,m}
                \frac{u^i}{i!}\frac{v^j}{j!}.
                \label{eq:generating-vector}
            \end{align}
            Under a spatial translation, this vector transforms as
            \begin{align}
                e^{(zM_1-\bar{z}M_{-1})/2}\mathcal{F}(u,v)
                =
                E(u,v)\mathcal F(u,v),
                \label{eq:generating-translation}
            \end{align}
            with
            \begin{align}
                E(u,v)
                &=
                \exp\!\left(mzu+m\bar{z} v-mzu^2v\right).
                \label{eq:E-function}
            \end{align}
            The translation matrix elements for the three dual bras are
            \begin{align}
                \bra*{\widetilde{0,0}}e^{(zM_1-\bar{z}M_{-1})/2}
                \ket{i,j}_{-+}
                &=
                i!j![u^iv^j]E(u,v),\nonumber\\
                \bra*{\widetilde{1,0}}e^{(zM_1-\bar{z}M_{-1})/2}
                \ket{i,j}_{-+}
                &=
                i!j![u^iv^j]uE(u,v),\nonumber\\
                (-\bra*{\widetilde{1,0}}L_1)e^{(zM_1-\bar{z}M_{-1})/2}
                \ket{i,j}_{-+}
                &=
                i!j![u^iv^j]u^2E(u,v).
                \label{eq:coefficient-extraction}
            \end{align}
            The coefficients of $E(u,v)$ are
            \begin{align}
                [u^i v^j]E(u,v)
                =
                \sum_{r=0}^{\min(\lfloor i/2\rfloor,j)}
                \frac{(-1)^r (mz)^{i-r}(m\bar{z})^{j-r}}
                {(i-2r)!(j-r)!r!}.
                \label{eq:E-coefficient}
            \end{align}

            \begin{align}
            H_p(s,z,\bar z)
            :=
            \sum_{k\geq\max(0,-p)}
            \frac{k!}{s^k}
            [u^{k+p}v^k]E(u,v),
            \qquad p=-2,-1,0,1,2.
            \end{align}
            Here $[u^av^b]$ extracts the coefficient of $u^av^b$.
            For $p=0,\pm1,\pm2$, the sums have the closed forms
            \begin{align}
                H_0
                &=\frac{\tau}{R}e^{m(\tau-R)},\label{eq:H0}\\
                H_{-1}
                &=-\frac{\tau\bar{z}}{R(\tau+R)}e^{m(\tau-R)},\label{eq:Hminus1}\\
                H_{-2}
                &=\frac{\tau\bar{z}^2}{R(\tau+R)^2}e^{m(\tau-R)},\label{eq:Hminus2}\\
                H_1
                &=\frac{\tau z}{R}\left(\frac{e^{m(\tau-R)}-1}{\tau-R}
                +\frac{1}{\tau+R}\right),\label{eq:H1}\\
                H_2
                &=\frac{\tau z^2}{R}\left(\frac{e^{m(\tau-R)}-1-m(\tau-R)}{(\tau-R)^2}
                +\frac{m}{\tau+R}+\frac{1}{(\tau+R)^2}\right).
                \label{eq:H2}
            \end{align}
            Here $\tau=-s/(2m)$ and $R=\sqrt{\tau^2+z\bar{z}}$.
            Derivatives with respect to $s$ below are taken at fixed $m,z,\bar{z}$.
            The apparent singularities at $R=\tau$ are removable.

            To evaluate the pairings with the states in eq.\,\eqref{eq:negative-reordered-states}, we use the differential operators
            \begin{align}
                \mathcal D&:=-s\frac{\partial}{\partial s},\nonumber\\
                \mathcal O_s&:=s^2-2s(\mathcal D+1)
                +(\mathcal D+1)(\mathcal D+2),\nonumber\\
                \mathcal P_s&:=(\mathcal D+1)(\mathcal D+2)
                -s(\mathcal D+1),\nonumber\\
                \mathcal Q_s&:=(\mathcal D+1)(\mathcal D+2).
                \label{eq:finite-differential-operators}
            \end{align}
            In the component ordering $(+1,0,-1)$, the nine pairings are
            \begin{align}
                I^{(-)}(X)
                =\mathcal{B}_0\frac{e^{-m\tau}}{\tau^3}
                \begin{pmatrix}
                    \mathcal{Q}_s H_0 & \sqrt{2}\,\mathcal{P}_s H_{-1} & \mathcal{O}_s H_{-2}\\
                    \sqrt{2}\,\mathcal{Q}_s H_1 & 2\mathcal{P}_s H_0 & \sqrt{2}\,\mathcal{O}_s H_{-1}\\
                    \mathcal{Q}_s H_2 & \sqrt{2}\,\mathcal{P}_s H_1 & \mathcal{O}_s H_0
                \end{pmatrix}.
                \label{eq:nine-I-pairings}
            \end{align}

            We write the result in terms of the three coefficients
            \begin{align}
                \mathsf A&:=1+\frac{1}{mR}+\frac{1}{m^2R^2},\nonumber\\
                \mathsf B&:=1+\frac{3}{mR}+\frac{3}{m^2R^2},\nonumber\\
                \mathsf C&:=1+\frac{1}{mR}.
                \label{eq:radial-ABC}
            \end{align}
            We now evaluate the differential operators in eq.\,\eqref{eq:nine-I-pairings} to write the matrix explicitly,
            \begin{align}
                I^{(-)}(X)
                &=
                \frac{m^2\mathcal B_0e^{-mR}}{R}
                \begin{pmatrix}
                    \displaystyle
                    2\mathsf A-\mathsf B\frac{\rho^2}{R^2}
                    -2\mathsf C\frac{\tau}{R}
                    &
                    \displaystyle
                    \sqrt2\,\bar z\frac{\mathsf C R-\mathsf B\tau}{R^2}
                    &
                    \displaystyle
                    \mathsf B\frac{\bar z^{\,2}}{R^2}
                    \\[4mm]
                    \displaystyle
                    \sqrt2\,z\frac{\mathsf B\tau-\mathsf C R}{R^2}
                    &
                    \displaystyle
                    2\left(-\mathsf A+\mathsf B\frac{\tau^2}{R^2}\right)
                    &
                    \displaystyle
                    -\sqrt2\,\bar z\frac{\mathsf B\tau+\mathsf C R}{R^2}
                    \\[4mm]
                    \displaystyle
                    \mathsf B\frac{z^2}{R^2}
                    &
                    \displaystyle
                    \sqrt2\,z\frac{\mathsf B\tau+\mathsf C R}{R^2}
                    &
                    \displaystyle
                    2\mathsf A-\mathsf B\frac{\rho^2}{R^2}
                    +2\mathsf C\frac{\tau}{R}
                \end{pmatrix}_{\!(+1,0,-1)} .
                \label{eq:I-closed-matrix}
            \end{align}

            We compare this matrix with the massive helicity propagators in Cartesian coordinates.
            With $\epsilon_{\tau xy}=+1$, the scalar Green function and the two helicity propagators are
            \begin{align}
                G_m^{\mathrm E}(R)
                &:=\frac{e^{-mR}}{4\pi R},
                \label{eq:scalar-green}\\
                G^{(-)}_{\mu\nu}(X)
                &:=
                \frac12
                \left(
                    \delta_{\mu\nu}
                    -\frac{1}{m^2}\partial_\mu\partial_\nu
                    +\frac{i}{m}\epsilon_{\mu\nu\rho}\partial_\rho
                \right)
                G_m^{\mathrm E}(R),
                \label{eq:helicity-minus-propagator}\\
                G^{(+)}_{\mu\nu}(X)
                &:=
                \frac12
                \left(
                    \delta_{\mu\nu}
                    -\frac{1}{m^2}\partial_\mu\partial_\nu
                    -\frac{i}{m}\epsilon_{\mu\nu\rho}\partial_\rho
                \right)
                G_m^{\mathrm E}(R).
                \label{eq:helicity-plus-propagator}
            \end{align}
            The derivatives of the scalar seed are
            \begin{align}
                \partial_\mu G_m^{\mathrm E}
                &=-m\mathsf C\,n_\mu G_m^{\mathrm E},\nonumber\\
                \partial_\mu\partial_\nu G_m^{\mathrm E}
                &=m^2G_m^{\mathrm E}
                \left[-(\mathsf A-1)\delta_{\mu\nu}
                +\mathsf B n_\mu n_\nu\right],
                \qquad
                n_\mu:=\frac{X_\mu}{R}.
                \label{eq:scalar-derivatives}
            \end{align}
            For the negative-helicity branch, eq.\,\eqref{eq:helicity-minus-propagator} therefore takes the form
            \begin{align}
                G^{(-)}_{\mu\nu}(X)
                =
                \frac{e^{-mR}}{8\pi R}
                \left[
                    \mathsf A\,\delta_{\mu\nu}
                    -\mathsf B\,n_\mu n_\nu
                    -i\mathsf C\,\epsilon_{\mu\nu\rho}n_\rho
                \right].
                \label{eq:Gminus-compact}
            \end{align}

            For the comparison with $I^{(-)}$, we order the Cartesian components as $(A_\tau,A_x,A_y)$ and the canonical helicity components as $(+1,0,-1)$,
            \begin{align}
                A_{+1}^{\mathrm{can}}
                &:=\frac{A_x+i A_y}{\sqrt2},
                &
                A_0^{\mathrm{can}}&:=A_\tau,
                &
                A_{-1}^{\mathrm{can}}
                &:=\frac{A_x-i A_y}{\sqrt2}.
                \label{eq:canonical-helicity-components}
            \end{align}
            The transformation between the two component bases is
            \begin{align}
                \bm A_H&=U\bm A_C,
                \qquad
                U=
                \begin{pmatrix}
                    0&1/\sqrt2&i/\sqrt2\\
                    1&0&0\\
                    0&1/\sqrt2&-i/\sqrt2
                \end{pmatrix},
                \qquad
                U^\dagger=
                \begin{pmatrix}
                    0&1&0\\
                    1/\sqrt2&0&1/\sqrt2\\
                    -i/\sqrt2&0&i/\sqrt2
                \end{pmatrix}.
                \label{eq:U-matrix}
            \end{align}
            In the helicity basis, the canonical propagator is
            \begin{align}
                G_H^{(-)}=UG_C^{(-)}U^\dagger.
                \label{eq:canonical-GH}
            \end{align}
            The pairing matrix and the canonical propagator are related by
            \begin{align}
                D_L&:=\operatorname{diag}(-1,1,1),
                &
                D_R&:=\operatorname{diag}(-1,-1,1),
                \label{eq:DL-DR}\\
                I^{(-)}&=D_LG_H^{(-)}D_R.
                \label{eq:I-G-dictionary}
            \end{align}
            In Cartesian components, this relation reads
            \begin{align}
                G_C^{(-)}=U^\dagger D_L I^{(-)}D_R U.
                \label{eq:Cartesian-reconstruction}
            \end{align}
            The transformations on the two indices differ because the bra belongs to the algebraic dual module rather than being the Hermitian conjugate of the ket.

            On the Euclidean time axis $X=(R,0,0)$, the pairing matrix reduces to
            \begin{align}
                I^{(-)}(R,0,0)
                =
                \frac{e^{-mR}}{8\pi m^2R^3}
                \operatorname{diag}\!\left(
                    1,
                    2(1+mR),
                    1+2mR+2m^2R^2
                \right).
                \label{eq:I-axis-final}
            \end{align}
            We fix the normalization by matching the propagator on this axis,
            \begin{align}
                \mathcal B_0=\frac{1}{16\pi m^2}.
                \label{eq:B0-final}
            \end{align}
            For this value of $\mathcal B_0$, the relation $I^{(-)}=D_LG_H^{(-)}D_R$ also holds at general $X$, so the pairings reproduce all components of the helicity propagator.

            The parity transformation in eq.~(3.32) acts on the spatial coordinates as $(x,y)\mapsto(-x,y)$.
            The pairing for the positive-helicity branch is therefore
            \begin{align}
            I^{(+)}(\tau,x,y)
            =
            P_H I^{(-)}(\tau,-x,y)P_H.
            \label{eq:I-plus-parity}
            \end{align}
            The parity-invariant Proca propagator is the sum of the two helicity propagators,
            \begin{align}
                G_{\mu\nu}^{\mathrm{Proca}}(X)
                &=
                G_{\mu\nu}^{(+)}(X)+G_{\mu\nu}^{(-)}(X)\nonumber\\
                &=
                \left(
                    \delta_{\mu\nu}
                    -\frac{1}{m^2}\partial_\mu\partial_\nu
                \right)
                G_m^{\mathrm E}(R).
                \label{eq:proca-flat-propagator}
            \end{align}
            The terms proportional to $\epsilon_{\mu\nu\rho}$ cancel between the two branches, as required by parity invariance.

        \subsection{Obstruction in the highest weight representation}
            In ref.\,\cite{Hao:2025btl}, it is shown that the bulk local state reconstruction fails for a massive scalar field when we use the highest weight representation (HWR) instead of the induced representation \eqref{eq:induced_rep}.
            Here, the HWR refers to the BMS HWR, not the Virasoro one.
            Ref.\,\cite{Hao:2025btl} therefore argues that the sensible representation for Carrollian holography is the induced representation rather than the HWR.
            In this subsection, we further ground this claim by showing the HWR also fails to reconstruct the massive spin-1 field.

            A BMS$_3$ highest weight primary $|\Delta,\xi\rangle_{\mathrm{HW}}$ satisfies
            \begin{align}
                \begin{aligned}
                    L_0|\Delta,\xi\rangle_{\mathrm{HW}}&=\Delta|\Delta,\xi\rangle_{\mathrm{HW}}, \\
                    M_0|\Delta,\xi\rangle_{\mathrm{HW}}&=\xi|\Delta,\xi\rangle_{\mathrm{HW}}, \\
                    L_n|\Delta,\xi\rangle_{\mathrm{HW}}&=0\quad(n>0), \\
                    M_n|\Delta,\xi\rangle_{\mathrm{HW}}&=0\quad(n>0).
                \end{aligned}
                \label{eq:carrollian-highest-weight-primary}
            \end{align}
            A descendant state is given by $L_{-1}^{i}M_{-1}^{j}|\Delta,\xi\rangle_{\mathrm{HW}}$, whose $L_0$ eigenvalue is $\Delta+N$ ($N:=i+j$).
            Since $L_0$ is diagonal in the level decomposition, components at different levels cannot cancel in an $L_0$ eigenstate.
            For a nontrivial solution, the $q=-1$ component must be nonzero.
            To prove this, if we assume $|\Phi_{-1}(t)\rangle=0$ and apply eq.\,\eqref{eq:intrinsic-B} with $q=-1$ and then $q=0$,
            \begin{align}
                |\Phi_0(t)\rangle&=-\frac{1}{\sqrt{2}}(L_{-1}+itM_{-1})|\Phi_{-1}(t)\rangle=0,\nonumber\\
                |\Phi_1(t)\rangle&=-\frac{1}{\sqrt{2}}(L_{-1}+itM_{-1})|\Phi_0(t)\rangle=0.
            \end{align}
            All three components would then vanish, contradicting the assumption of a nontrivial solution.
            For this nonzero component, the local covariance condition in eq.\,\eqref{eq:intrinsic-L0} reads
            \begin{align}
                (L_0+1)|\Phi_{-1}(t)\rangle=0.
            \end{align}
            On a descendant of level $N$, the operator $L_0+1$ acts by multiplication by $\Delta+N+1$.
            Since different levels belong to distinct $L_0$ eigenspaces, this equation must hold separately at each level in the expansion of $|\Phi_{-1}(t)\rangle$.
            Every level with $\Delta+N+1\neq0$ must have zero coefficient.
            A nonzero $|\Phi_{-1}(t)\rangle$ can thus have support only at a level $N_{-1}$ satisfying
            \begin{align}
                \Delta+N_{-1}=-1,
                \qquad N_{-1}\in\mathbb{Z}_{\geq0},
                \qquad \Delta\in\mathbb{Z}_{\leq-1}.
                \label{eq:carrollian-highest-weight-spectrum-bound}
            \end{align}

            The mass and helicity impose a conflicting condition on $\Delta$.
            The global BMS$_3$ Casimirs are
            \begin{align}
                C_1&=M_0^2-M_{-1}M_1,\nonumber\\
                C_2&=2L_0M_0-\frac{1}{2}\left(
                    L_{-1}M_1+L_1M_{-1}+M_1L_{-1}+M_{-1}L_1
                \right).
                \label{eq:carrollian-global-casimirs}
            \end{align}
            On the primary in eq.\,\eqref{eq:carrollian-highest-weight-primary}, their eigenvalues are
            \footnote{
                By contrast, the induced primary used above has eigenvalues
                \begin{align}
                    C_1|\Delta,\xi\rangle&=\xi^2|\Delta,\xi\rangle,\nonumber\\
                    C_2|\Delta,\xi\rangle&=2\Delta\xi|\Delta,\xi\rangle.
                    \label{eq:carrollian-induced-casimirs}
                \end{align}
                The induced solutions with $\xi=m$ and $\Delta=\lambda=\pm1$ satisfy this requirement by eq.\,\eqref{eq:carrollian-induced-casimirs}.
            }
            \begin{align}
                \begin{aligned}
                    C_1|\Delta,\xi\rangle_{\mathrm{HW}}&=\xi^2|\Delta,\xi\rangle_{\mathrm{HW}}, \\
                    C_2|\Delta,\xi\rangle_{\mathrm{HW}}&=2(\Delta-1)\xi|\Delta,\xi\rangle_{\mathrm{HW}}.
                \end{aligned}
                \label{eq:carrollian-highest-weight-casimirs}
            \end{align}
            The shift by $-1$ in the second eigenvalue comes from $M_1L_{-1}|\Delta,\xi\rangle_{\mathrm{HW}}=L_1M_{-1}|\Delta,\xi\rangle_{\mathrm{HW}}=2\xi|\Delta,\xi\rangle_{\mathrm{HW}}$.
            Both Casimirs commute with the global generators, so every state in this sector has the eigenvalues in eq.\,\eqref{eq:carrollian-highest-weight-casimirs}.
            A bulk one-particle state of mass $m$ and helicity $\lambda=\pm1$ has the Poincar\'e Casimir eigenvalues
            \begin{align}
                C_1=m^2,
                \qquad
                C_2=2\lambda m.
                \label{eq:carrollian-spin-one-casimirs}
            \end{align}
            A reconstruction consistent with Poincar\'e symmetry must preserve these eigenvalues, so the boundary state representing this bulk excitation must satisfy eq.\,\eqref{eq:carrollian-spin-one-casimirs}.
            For a highest weight module, eq.\,\eqref{eq:carrollian-highest-weight-casimirs} requires $\xi^2=m^2$ and $2(\Delta-1)\xi=2\lambda m$.
            For $\xi=m\neq0$, the second condition fixes $\Delta=1+\lambda$.
            The other root $\xi=-m$ instead requires $\Delta=1-\lambda$.
            Either choice restricts $\Delta$ to $0$ or $2$.
            Neither value satisfies eq.\,\eqref{eq:carrollian-highest-weight-spectrum-bound}.
            The only state in this sector that obeys both the local covariance conditions and the massive spin-1 Casimir conditions is therefore the zero state.

            The level argument uses the diagonal action
            \begin{align}
                L_0L_{-1}^{i}M_{-1}^{j}|\Delta,\xi\rangle_{\mathrm{HW}}=(\Delta+i+j)L_{-1}^{i}M_{-1}^{j}|\Delta,\xi\rangle_{\mathrm{HW}},
            \end{align}
            in the highest weight module considered here.
            In the $m=0$ limit of eq.\,\eqref{eq:carrollian-spin-one-casimirs}, the Casimir conditions become
            \begin{align}
                \xi^2=0,\qquad 2(\Delta-1)\xi=0.
            \end{align}
            The first condition fixes $\xi=0$, after which the second is $0=0$ for every $\Delta$.
            Unlike the massive case, these conditions do not restrict $\Delta$ to $0$ or $2$ and hence do not contradict $\Delta=-1-N_{-1}$.
            The Casimir comparison alone therefore establishes neither the existence nor the absence of a massless reconstruction.

    \section{Flat limit from AdS/CFT}
        \label{sec:flat-limit}
        This section relates the AdS and Carrollian bulk local states constructed in Sections \ref{sec:ads} and \ref{sec:intrinsic}, and therefore presupposes both constructions.
        Following the scalar analysis in ref.\,\cite{Hao:2025btl}, we compare the symmetry representations, their Casimirs, the local states, and finally their two-point functions.
        The physical mass $m>0$ is fixed throughout the limit.

        \subsection{Contraction of the algebra and the representation}
            We begin by summarizing the Virasoro algebra $\operatorname{Vir}\times\overline{\operatorname{Vir}}$ with $c_{\text{AdS}_3}=\overline{c}_{\text{AdS}_3}=\frac{3\ell}{2G}$, the Brown--Henneaux value for AdS$_3$/CFT$_2$ with AdS radius $\ell$,
            \begin{align}
                \begin{aligned}
                    [\mathcal{L}_m,\mathcal{L}_n]&=(m-n)\mathcal{L}_{m+n}+\frac{c_{\text{AdS}_3}}{12}(m^3-m)\delta_{m+n,0}, \\
                    [\overline{\mathcal{L}}_m,\overline{\mathcal{L}}_n]&=(m-n)\overline{\mathcal{L}}_{m+n}+\frac{\overline{c}_{\text{AdS}_3}}{12}(m^3-m)\delta_{m+n,0}, \\
                    [\mathcal{L}_m,\overline{\mathcal{L}}_n]&=0,
                \end{aligned}
                \qquad(m,n\in\mathbb{Z}).
                \label{eq: Virasoro commutation relations}
            \end{align}
            We adopt the calligraphy $\mathcal{L}_n$ to denote the Virasoro generators for distinction from BMS$_3$ generators $L_n$.
            The massive AdS representation is generated from a primary state $|h,\overline{h}\rangle:=|h\rangle\otimes|\overline{h}\rangle$ satisfying
            \begin{align}
                \mathcal{L}_0|h,\overline{h}\rangle=h|h,\overline{h}\rangle,\:\:\overline{\mathcal{L}}_0|h,\overline{h}\rangle=\overline{h}\:|h,\overline{h}\rangle,\:\mathcal{L}_n|h,\overline{h}\rangle=\overline{\mathcal{L}}_n|h,\overline{h}\rangle=0\:(n>0). \label{eq: highest weight condition}
            \end{align}
            The corresponding highest weight module is spanned by the descendants
            \begin{align}
                \mathcal{L}_{-n_1}\cdots\mathcal{L}_{-n_k}\overline{\mathcal{L}}_{-\overline{n}_1}\cdots\overline{\mathcal{L}}_{-\overline{n}_l}|h,\overline{h}\rangle\quad(n_1\geq \cdots\geq n_k\geq 0,\,\overline{n}_1\geq \cdots\geq \overline{n}_l\geq 0),
            \end{align}
            with the standard adjoint
            \begin{align}
                \mathcal{L}_n^{\dagger}=\mathcal{L}_{-n},\:\overline{\mathcal{L}}_n^{\dagger}=\overline{\mathcal{L}}_{-n}. \label{eq: Virasoro conjugation rule}
            \end{align}
            Then, let us apply a suitable procedure to reproduce the BMS$_3$ algebra from the AdS$_3$ Virasoro algebra \eqref{eq: Virasoro commutation relations} in its flat limit $\ell\to\infty$.
            We use the contraction of ref.\,\cite{Hao:2025btl},
            \begin{align}
                L_{n}(\ell):=\mathcal{L}_n-(-1)^n\overline{ \mathcal{L}}_{-n},\;\;
                M_{n}(\ell):=\frac{1}{\ell}(\mathcal{L}_n+(-1)^n\overline{ \mathcal{L}}_{-n}),   \label{eq: generator relation between flat and AdS}
            \end{align}
            which obey the finite-radius commutators
            \begin{align}
                \begin{aligned}
                    [L_m(\ell),L_n(\ell)] & =(m-n)L_{m+n}(\ell)+\frac{c_L}{12}(m^3-m)\delta_{m+n,0}, \\
                    [L_m(\ell),M_n(\ell)] & =(m-n)M_{m+n}(\ell)+\frac{c_M}{12}(m^3-m)\delta_{m+n,0}, \\
                    [M_m(\ell),M_n(\ell)] & =\frac{1}{\ell^2}\Big((m-n)L_{m+n}(\ell)+\frac{c_L}{12}(m^3-m)\delta_{m+n,0}\Big).
                \end{aligned} \label{eq: BMS-like commutators before the flat limit}
            \end{align}
            The central terms $c_L$, $c_M$ are related to the Virasoro central charges $c_{\text{AdS}_3}$, $\overline{c}_{\text{AdS}_3}$ by
            \begin{align}
                c_L:=c_{\text{AdS}_3}-\overline{c}_{\text{AdS}_3}=0,\quad c_M:=\frac{c_{\text{AdS}_3}+\overline{c}_{\text{AdS}_3}}{\ell}=\frac{3}{G},  \label{eq: BMS-like central charge}
            \end{align}
            which are the standard BMS$_3$ central charges of three-dimensional Einstein gravity.
            It also follows from eq.\,\eqref{eq: Virasoro conjugation rule} that eq.\,\eqref{eq: generator relation between flat and AdS} properly satisfies the conjugation rule
            \begin{align}
                L_n(\ell)^{\dagger}=L_{-n}(\ell),\;\; M_n(\ell)^{\dagger}=M_{-n}(\ell). \label{eq: flat conjugation before flat limit}
            \end{align}
            By taking $\ell\to\infty$, the commutators \eqref{eq: BMS-like commutators before the flat limit} reduce to the BMS$_3$ algebra with the central charges in \eqref{eq: BMS-like central charge}.
            The highest weight conditions also have a direct expression in these generators.
            We rewrite \eqref{eq: highest weight condition} in terms of the new generators \eqref{eq: generator relation between flat and AdS}
            \begin{gather}
                L_0(\ell)|h,\overline{h}\rangle=(h-\overline{h})|h,\overline{h}\rangle,\:\:M_0(\ell)|h,\overline{h}\rangle=\frac{h+\overline{h}}{\ell}|h,\overline{h}\rangle, \\
                \Big(M_n(\ell)+\frac{1}{\ell}L_n(\ell)\Big)|h,\overline{h}\rangle=0,\:\:\Big(M_{-n}(\ell)-\frac{1}{\ell}L_{-n}(\ell)\Big)|h,\overline{h}\rangle=0\quad(n>0). \label{eq: highest weight condition in AdS represented in flat generators}
            \end{gather}
            On a family of modules with $\xi=\lim_{\ell\to\infty}(h+\overline{h})/\ell$ and fixed $h-\overline{h}=\lambda$, these conditions result in the induced conditions
            \begin{align}
                M_0|\lambda,\xi\rangle=\xi|\lambda,\xi\rangle,\:M_n|\lambda,\xi\rangle=0\,(n\neq 0),\:L_0|\lambda,\xi\rangle=\lambda|\lambda,\xi\rangle,  \label{eq: BMS induced from AdS highest}
            \end{align}
            where $\displaystyle L_n:=\lim_{\ell\to\infty}L_n(\ell),\;M_n:=\lim_{\ell\to\infty}M_n(\ell),\;|\lambda,\xi\rangle:=\lim_{\ell\to\infty}|h,\overline{h}\rangle$.
            Here the primary and each fixed ordered descendant are identified algebraically along the family of modules.
            For the vector branches, eqs.\,\eqref{eq:ads-vector-weights} and \eqref{eq:ads-vector-mass} give
            \begin{align}
                h_{\lambda}=\frac{1+m\ell+\lambda}{2},\qquad
                \bar{h}_{\lambda}=\frac{1+m\ell-\lambda}{2},\qquad
                \xi=m,\qquad \Delta_{\mathrm{flat}}=\lambda=\pm1.
                \label{eq:ads-to-flat-spin-one-weights}
            \end{align}
            Thus both induced primaries in eq.\,\eqref{eq:two-flat-branches} arise from massive AdS representations.
            We use $\Delta_{\mathrm{AdS}}=h+\bar{h}$ below to distinguish the diverging AdS energy from the finite Carrollian rotation weight.

        \subsection{Mass and spin from the Casimirs}
            The contraction must preserve both the mass and the helicity of the excitation.
            We use the global conformal Casimirs
            \begin{align}
                \mathcal{C}&=\mathcal{L}_0^2-\frac{1}{2}\{\mathcal{L}_1,\mathcal{L}_{-1}\},
                &
                \overline{\mathcal{C}}
                &=\overline{\mathcal{L}}_0^2-\frac{1}{2}\{\overline{\mathcal{L}}_1,\overline{\mathcal{L}}_{-1}\},
                \label{eq:flat-ads-casimirs}
            \end{align}
            with eigenvalues $h(h-1)$ and $\bar{h}(\bar{h}-1)$.
            Here $\{A,B\}=AB+BA$.
            Substitution of eq.\,\eqref{eq: generator relation between flat and AdS} gives the exact identities
            \begin{align}
                \frac{2}{\ell^2}(\mathcal{C}+\overline{\mathcal{C}})
                &=M_0(\ell)^2-\frac{1}{2}\{M_1(\ell),M_{-1}(\ell)\}
                +\frac{1}{\ell^2}\left(L_0(\ell)^2-\frac{1}{2}\{L_1(\ell),L_{-1}(\ell)\}\right),
                \label{eq:flat-casimir-mass}\\
                \frac{2}{\ell}(\mathcal{C}-\overline{\mathcal{C}})
                &=2L_0(\ell)M_0(\ell)
                -\frac{1}{2}\left(\{L_1(\ell),M_{-1}(\ell)\}+\{L_{-1}(\ell),M_1(\ell)\}\right).
                \label{eq:flat-casimir-spin}
            \end{align}
            On each fixed descendant, their limits are $C_1$ and $C_2$ in eq.\,\eqref{eq:carrollian-global-casimirs}.
            For $\Delta_{\mathrm{AdS}}=1+m\ell$ and $\lambda=\pm1$, their eigenvalues are
            \begin{align}
                \frac{2}{\ell^2}\big(h(h-1)+\bar{h}(\bar{h}-1)\big)
                &=\frac{\Delta_{\mathrm{AdS}}(\Delta_{\mathrm{AdS}}-2)+\lambda^2}{\ell^2}=m^2,
                \\
                \frac{2}{\ell}\big(h(h-1)-\bar{h}(\bar{h}-1)\big)
                &=\frac{2\lambda(\Delta_{\mathrm{AdS}}-1)}{\ell}=2m\lambda.
                \label{eq:flat-vector-casimir-values}
            \end{align}
            The contracted module therefore has the massive spin-1 Casimirs, not merely the same primary labels.

        \subsection{Bulk local state correspondence}
            We first compare the covariance conditions and then compute the descendant coefficients of the limiting state.
            Physical global coordinates are held fixed as $\ell\to\infty$,
            \begin{align}
                ds^2=-\left(1+\frac{r^2}{\ell^2}\right)dt^2
                +\frac{dr^2}{1+r^2/\ell^2}+r^2d\phi^2
                \longrightarrow-dt^2+dr^2+r^2d\phi^2.
                \label{eq:flat-global-metric}
            \end{align}
            In particular, the AdS center at physical time $t$ approaches the flat-space point $(t,0,0)$.
            With the time convention of eq.\,\eqref{eq:local-time-evolution}, its state is
            \begin{align}
                |\Phi_q^{(\lambda)}(t)\rangle_{\mathrm{AdS}}
                =e^{it(\mathcal{L}_0+\overline{\mathcal{L}}_0)/\ell}
                |\Phi_q^{(\lambda)}\rangle_{\mathrm{AdS}}.
                \label{eq:flat-ads-time-state}
            \end{align}
            We retain the primary phase, since it becomes the factor $e^{imt}$ in the flat state.
            The translated stabilizers are
            \begin{align}
                e^{\mp it/\ell}\mathcal{L}_{\pm1}+e^{\pm it/\ell}\overline{\mathcal{L}}_{\mp1}
                &=\cos(t/\ell)L_{\pm1}(\ell)\mp i\ell\sin(t/\ell)M_{\pm1}(\ell)
                \nonumber\\
                &\longrightarrow L_{\pm1}\mp itM_{\pm1}.
                \label{eq:flat-stabilizer-limit}
            \end{align}
            Thus the three AdS covariance equations become eqs.\,\eqref{eq:intrinsic-L0}--\eqref{eq:intrinsic-B}, including the mixing of vector components.

            To calculate the states themselves, we use the finite-radius descendant identity
            \begin{align}
                \mathcal{L}_{-1}^{p}\overline{\mathcal{L}}_{-1}^{q}|h,\overline{h}\rangle
                =\sum_{b=\max(0,q-p)}^{q}
                C_{p,q}(b)\,
                L_1(\ell)^bL_{-1}(\ell)^{p-q+b}|h,\overline{h}\rangle,
                \label{ea:adscft-to-flat-descendant}
            \end{align}
            where
            \begin{align}
                C_{p,q}(b)
                &=(-1)^{q-b}\binom{q}{b}
                \prod_{j=p-q+b+1}^{p}j(2h+j-1)\nonumber\\
                &=\frac{(-1)^{q-b}\Gamma(q+1)\Gamma(p+1)\Gamma(p+2h)}
                {\Gamma(b+1)\Gamma(q-b+1)\Gamma(p-q+b+1)\Gamma(p-q+b+2h)}.
                \label{eq:ads-descendant-conversion-coefficient}
            \end{align}
            The product is one when its lower limit exceeds its upper limit.
            If individual gamma functions have poles, the ratio in the second line is understood through this finite-product expression.
            See Appendix \ref{appendix:ads-descendant-conversion} for proof.

            Interchanging the holomorphic and antiholomorphic modules gives the ordering needed for the positive-helicity branch,
            \begin{align}
                \mathcal{L}_{-1}^{k}\overline{\mathcal{L}}_{-1}^{k+r}|h,\bar{h}\rangle_+
                &=\sum_{a=0}^{k}
                \frac{(-1)^{k-a}k!(k+r)!\,\Gamma(k+r+m\ell)}
                {a!(k-a)!(a+r)!\,\Gamma(a+r+m\ell)}
                L_{-1}(\ell)^aL_1(\ell)^{a+r}|h,\bar{h}\rangle_+,
                \label{eq:flat-descendant-positive}
            \end{align}
            where $r=0,1,2$.
            The normalized descendants in Section \ref{sec:ads} cancel the square roots in $a_k,b_k,c_k$.
            Consequently the positive-helicity triplet takes the unnormalized form
            \begin{align}
                |\Phi_{1-r}^{(+)}(t)\rangle_{\mathrm{AdS}}
                &=\mathcal{N}_A(-1)^r\sqrt{\binom{2}{r}}\,
                e^{i(\Delta_{\mathrm{AdS}}+r)t/\ell}
                \sum_{k=0}^{\infty}
                \frac{(-1)^k z^k}{k!(m\ell)_{k+r}}
                \mathcal{L}_{-1}^{k}\overline{\mathcal{L}}_{-1}^{k+r}|h,\bar{h}\rangle_+,
                \qquad z=e^{2it/\ell}.
                \label{eq:flat-ads-unnormalized}
            \end{align}
            We initially take $\operatorname{Im}t>0$, so that $|z|<1$.
            For any fixed ordered descendant, the sum over the original AdS level then converges absolutely.
            Equations \eqref{eq:flat-descendant-positive} and \eqref{eq:flat-ads-unnormalized} reduce that sum to
            \begin{align}
                \sum_{k=a}^{\infty}\frac{(k+r)!}{(k-a)!}z^k
                &=\frac{(a+r)!\,z^a}{(1-z)^{a+r+1}},
                \label{eq:flat-binomial-sum}\\
                |\Phi_{1-r}^{(+)}(t)\rangle_{\mathrm{AdS}}
                &=\mathcal{N}_A\sum_{a=0}^{\infty}f_{a,r}(\ell,t)
                L_{-1}(\ell)^aL_1(\ell)^{a+r}|h,\bar{h}\rangle_+,
                \\
                f_{a,r}(\ell,t)
                &=\frac{(-1)^{a+r}\sqrt{\binom{2}{r}}\,
                e^{i(\Delta_{\mathrm{AdS}}+r)t/\ell}z^a}
                {a!(m\ell)_{a+r}(1-z)^{a+r+1}}.
                \label{eq:flat-exact-coefficient}
            \end{align}
            The binomial sum follows by differentiating the geometric series $a+r$ times and multiplying by $z^a$.
            It includes arbitrarily high AdS levels before any limit is taken.

            For fixed $a,r,t$, the factors controlling the limit are
            \begin{align}
                (m\ell)_{a+r}&=(m\ell)^{a+r}\big[1+O(\ell^{-1})\big],
                &
                1-z&=-\frac{2it}{\ell}+O(\ell^{-2}),
                &
                e^{i\Delta_{\mathrm{AdS}}t/\ell}&\longrightarrow e^{imt}.
            \end{align}
            We choose the remaining common normalization so that the scalar seed in the two-point function is $e^{-mR}/(4\pi R)$, as verified below in eq.\,\eqref{eq:flat-propagator-axis-limit}.
            The corresponding coefficient limit is
            \begin{align}
                \lim_{\ell\to\infty}\frac{f_{a,r}(\ell,t)}{2\pi \ell}
                =\frac{i e^{imt}}{4\pi t}
                \sqrt{\binom{2}{r}}\,
                \frac{\alpha(t)^{a+r}}{a!},
                \qquad \alpha(t)=-\frac{i}{2mt}.
                \label{eq:flat-coefficient-limit}
            \end{align}
            Therefore
            \begin{align}
                \lim_{\ell\to\infty}\frac{|\Phi_{1-r}^{(+)}(t)\rangle_{\mathrm{AdS}}}
                {2\pi \ell\mathcal{N}_A}
                &=\frac{i e^{imt}}{4\pi t}
                \sqrt{\binom{2}{r}}\,\alpha(t)^r
                \sum_{a=0}^{\infty}\frac{\alpha(t)^a}{a!}
                |a,a+r\rangle_{-+}^{1,m}.
                \label{eq:flat-state-limit-positive}
            \end{align}
            This is eq.\,\eqref{eq:flat-compact} with $A_+=i/(4\pi)$.
            For the negative-helicity branch, we interchange the two conformal modules in the calculation.
            Its AdS coefficients have $\sqrt{\binom{2}{r}}$ in place of $(-1)^r\sqrt{\binom{2}{r}}$, and its ordered descendants are $L_1(\ell)^aL_{-1}(\ell)^{a+r}$.
            The result is
            \begin{align}
                \lim_{\ell\to\infty}\frac{|\Phi_{-1+r}^{(-)}(t)\rangle_{\mathrm{AdS}}}
                {2\pi \ell\mathcal{N}_A}
                &=\frac{i e^{imt}}{4\pi t}
                \sqrt{\binom{2}{r}}\,[-\alpha(t)]^r
                \sum_{a=0}^{\infty}\frac{\alpha(t)^a}{a!}
                |a,a+r\rangle_{+-}^{-1,m},
                \label{eq:flat-state-limit-negative}
            \end{align}
            which fixes $A_-=A_+$ and reproduces eq.\,\eqref{eq:flat-compact-minus}.
            These limits mean convergence of each coefficient in the indicated ordered basis, or equivalently of pairings with functionals of finite support in that basis.
            No exchange of the large-$\ell$ limit with the original AdS level sum is required.
            Lorentzian expressions are the boundary values $t\to t+i0$; the origin itself remains a singular local state.

            The same calculation also identifies the dual used in Section \ref{subsec:carrollian-dual-basis}.
            For the negative branch, the AdS bras at the origin obey the right-action equations \eqref{eq:dual-right-actions}, with $L_{\pm1}$ replaced by $L_{\pm1}(\ell)$.
            Their overlaps with the primary are $\mathcal{N}_A^*$ for $q=-1$ and zero for $q=0,+1$.
            Repeated right action on the ordered basis thus gives, already at finite radius,
            \begin{align}
                \frac{{}_{\mathrm{AdS}}\langle\Phi_{-1}^{(-)}|}
                {\mathcal{N}_A^*}L_{-1}(\ell)^iL_1(\ell)^j|h,\bar{h}\rangle_-
                &=\delta_{i0}\delta_{j0},
                \\
                \frac{{}_{\mathrm{AdS}}\langle\Phi_{0}^{(-)}|}
                {\mathcal{N}_A^*}L_{-1}(\ell)^iL_1(\ell)^j|h,\bar{h}\rangle_-
                &=\sqrt{2}\delta_{i1}\delta_{j0},
                \\
                \frac{{}_{\mathrm{AdS}}\langle\Phi_{1}^{(-)}|}
                {\mathcal{N}_A^*}L_{-1}(\ell)^iL_1(\ell)^j|h,\bar{h}\rangle_-
                &=2\delta_{i2}\delta_{j0}.
                \label{eq:flat-dual-contraction}
            \end{align}
            These are precisely $\bra*{\widetilde{0,0}}$, $\sqrt{2}\bra*{\widetilde{1,0}}$, and $-\bra*{\widetilde{1,0}}L_1$.
            The bra and ket therefore have different normalization factors in this algebraic contraction.

            For comparison with the explicit Euclidean states, the opposite ordering can be taken before the flat limit.
            In the contracted algebra, normal ordering uses
            \begin{align}
                L_1^bL_{-1}^A|\Delta,m\rangle
                =\sum_{s=0}^{\min(A,b)}
                \binom{b}{s}\frac{A!}{(A-s)!}
                (2\Delta+A-b)_s
                L_{-1}^{A-s}L_1^{b-s}|\Delta,m\rangle.
                \label{eq:flat-normal-order}
            \end{align}
            For $\Delta=-1$ and $A=b+r$, the factors $(-2+r)_s$ terminate at $s=2-r$.
            Thus only finitely many terms contribute to any coefficient in the reordered series.
            With $t=i\tau$ and $s=-2m\tau=\alpha(i\tau)^{-1}$, eq.\,\eqref{eq:flat-state-limit-negative} becomes eq.\,\eqref{eq:negative-reordered-states}, with
            \begin{align}
                \mathcal{N}_-(i\tau)\alpha(i\tau)^2
                =\frac{e^{-m\tau}}{16\pi m^2\tau^3}.
                \label{eq:flat-reordered-normalization}
            \end{align}
            This accounts for the extra $\tau^{-2}$ in the reordered prefactor and fixes $\mathcal{B}_0$ as in eq.\,\eqref{eq:B0-final}.

        \subsection{Two-point function correspondence}
            The bra and ket normalizations in the preceding subsection fix the normalization of the limiting overlap.
            We evaluate it first at a nonzero Euclidean separation $R$ along the time axis.
            There the geodesic distance is $R$ and $\chi=R/\ell$, so
            \begin{align}
                \Delta_{\mathrm{AdS}}&=1+m\ell,
                &
                e^{-(\Delta_{\mathrm{AdS}}-1)\chi}&=e^{-mR},
                &
                \Delta_{\mathrm{AdS}}(\Delta_{\mathrm{AdS}}-1)\sinh^3\chi
                &=\frac{m^2R^3}{\ell}\big[1+O(\ell^{-1})\big].
                \label{eq:flat-propagator-scaling}
            \end{align}
            Substitution into the three closed sums \eqref{eq:ads-Gpp}--\eqref{eq:ads-Gmm}, with the helicities reversed for the negative branch, gives
            \begin{align}
                \lim_{\ell\to\infty}\frac{G_{\mathrm{AdS}}^{(-)}(R/\ell)}
                {2\pi \ell|\mathcal{N}_A|^2}
                &=\frac{e^{-mR}}{8\pi m^2R^3}
                \operatorname{diag}\left(1,\;2(1+mR),\;1+2mR+2m^2R^2\right)
                \nonumber\\
                &=I^{(-)}(R,0,0).
                \label{eq:flat-propagator-axis-limit}
            \end{align}
            All three entries agree with eq.\,\eqref{eq:I-axis-final}, with a single common normalization.
            The positive branch follows by the parity relation \eqref{eq:I-plus-parity}.

            To extend this comparison away from the axis, we keep both physical points fixed in the coordinates of eq.\,\eqref{eq:flat-global-metric}.
            In Euclidean signature their AdS geodesic distance satisfies
            \begin{align}
                \cosh\frac{D_\ell}{\ell}
                &=\sqrt{1+\frac{r^2}{\ell^2}}\sqrt{1+\frac{r'^2}{\ell^2}}\,
                \cosh\frac{\tau-\tau'}{\ell}
                -\frac{rr'}{\ell^2}\cos(\phi-\phi')
                \nonumber\\
                &=1+\frac{R^2}{2\ell^2}+O(\ell^{-4}),
                \\
                R^2&=(\tau-\tau')^2+r^2+r'^2-2rr'\cos(\phi-\phi').
                \label{eq:flat-geodesic-limit}
            \end{align}
            Hence $D_\ell\to R$, while parallel transport in a smooth orthonormal frame approaches ordinary flat-space transport.
            Maximal symmetry determines the vector two-point function from its three components in a frame adapted to the connecting geodesic.
            The axis limit therefore fixes the tensor at every noncoincident pair of points.
            Applying the left and right component maps of eq.\,\eqref{eq:Cartesian-reconstruction}, and their parity counterparts, yields
            \begin{align}
                G_{\mu\nu}^{(\lambda)}(X-X')
                =\frac{1}{2}\left(
                \delta_{\mu\nu}-\frac{1}{m^2}\partial_\mu\partial_\nu
                -\frac{i\lambda}{m}\epsilon_{\mu\nu\rho}\partial_\rho
                \right)\frac{e^{-mR}}{4\pi R},
                \qquad R>0.
                \label{eq:flat-helicity-green-limit}
            \end{align}
            The sum over $\lambda=\pm1$ is the Proca two-point function in eq.\,\eqref{eq:proca-flat-propagator}.
            Lorentzian boundary values follow with the same $i0$ prescription as for the states.
            The comparison concerns separated-point two-point functions; contact terms in a sourced Green function require the corresponding distributional prescription.

            The fixed physical scales used here also enter the flat limit of the information metric in Section \ref{sec:spinone-info-flat}.
            There the regulator is held at $\tau_c=\ell\varepsilon$, and the relevant AdS levels grow as $k=O(\ell)$.
            The exact sum \eqref{eq:flat-binomial-sum} and the closed propagators above retain those levels, rather than approximating the state by a fixed finite set of AdS descendants.

%%%%%%%%%%%%%%%%%%%%%%%%%%%%%%%%%%%%%%%%%%%%%%%%%%%%%%%%
%%%%%%%%%%%%%%%%%%%%%%%%%%%%%%%%%%%%%%%%%%%%%%%%%%%%%%%%
\section{Massless Vector Case}
\label{app:massless}
%%%%%%%%%%%%%%%%%%%%%%%%%%%%%%%%%%%%%%%%%%%%%%%%%%%%%%%%
%%%%%%%%%%%%%%%%%%%%%%%%%%%%%%%%%%%%%%%%%%%%%%%%%%%%%%%%
So far we have been discussing massive vector fields. In this section we discuss massless vector fields.
There are two possible massless vector theories in AdS$_3$: Maxwell theory and Chern-Simons theory.
Due to its gauge symmetry, the former has one propagating bulk degree of freedom, while the latter has no propagating bulk degrees of freedom, but contains boundary degrees of freedom.

Before discussing each theory, let us consider representational aspects here.
We note that the coefficient $a_k$ in (\ref{eq:ads-ak}) diverges in the massless limit (i.e. $\Delta \to 1$).
This divergence of $a_k$ is not a physical divergence of a massless field, but it signals representation shortening and the appearance of gauge redundancy.
In the massless limit, we have
	\begin{align}
		(h_{+1}, \bar{h}_{+1}) \, \to \, (1,0) \qquad \textrm{or} \qquad (h_{-1}, \bar{h}_{-1}) \, \to \, (0,1) \, .
	\end{align}
Therefore, in this limit, we observe that some of the descendants become null states. For example,
	\begin{align}
		\Vert \ol{\mathcal{L}}_{-1}|1,0 \rangle \Vert^2\, = \, 2\bar{h}_{+1} \, = 0 \, , \qquad
        \Vert \mathcal{L}_{-1}|0,1 \rangle \Vert^2\, = \, 2h_{-1} \, = 0 \, .
	\end{align}
Hence, the normalized descendant basis used in Section~\ref{sec:ads} becomes singular due to these null states.
The $(1,0)$ and $(0,1)$ sectors describe holomorphic and anti-holomorphic conserved currents, and they are naturally contained in the Chern-Simons/topological sector.
The propagating Maxwell mode is instead described by the self-dual scalar sector $(h,\bar h)=(1,1)$.

%%%%%%%%%%%%%%%%%%%%%%%%%%%%%%%%%%%%%%%%%%%%%%%%%%%%%%%%
\subsection{Maxwell theory}
\label{sec:maxwell}
%%%%%%%%%%%%%%%%%%%%%%%%%%%%%%%%%%%%%%%%%%%%%%%%%%%%%%%%
In this subsection, we consider the Maxwell theory in AdS$_3$:
	\begin{align}
		I_{\textrm{MW}} \, = \, - \frac{1}{4g^2} \int_{\mathcal{M}} d^3x \sqrt{-g} \, F^{\mu\nu}F_{\mu\nu} \, + \, \frac{1}{g^2} \int_{\partial\mathcal{M}} \sqrt{-h} \, n_\mu A_\nu F^{\mu\nu} \, .
	\label{eq:I_MW}
	\end{align}
The variation of the gauge field on-shell leads to
	\begin{align}
		\delta_A I_{\textrm{MW}} \, = \, \frac{1}{g^2} \int_{\partial\mathcal{M}} \sqrt{-h} \, A_I \delta \mathcal{F}^I \, ,
	\end{align}
where $I$ is the boundary index and
	\begin{align}
		\mathcal{F}^\nu \, : = \, n_\mu F^{\mu\nu} \, .
	\end{align}
Therefore, the boundary term in (\ref{eq:I_MW}) corresponds to fixing $\mathcal{F}^I$ on the boundary.
This boundary condition leads to one propagating degree of freedom in the bulk of AdS$_3$ \cite{Marolf:2006nd}.
The other boundary condition one can think of is fixing $A^I$ on the boundary; however, this boundary condition leads to no propagating degrees of freedom in the bulk of AdS$_3$ \cite{Marolf:2006nd}.

The Maxwell equation and the Bianchi identity are given by
	\begin{align}
		d * F \, = \, 0 \, , \qquad dF \, = \, 0 \, .
	\end{align}
Since $*F$ is a one-form in three dimensions, the Maxwell equation implies locally
	\begin{align}
		* F \, = \, d\varphi \, , \quad \Leftrightarrow \quad F \, = \, *d\varphi \, ,
	\end{align}
with a scalar field $\varphi$.
Then the Bianchi identity leads to
	\begin{align}
		dF \, = \, d*d\varphi \, = \, 0 \, , \quad \Leftrightarrow \quad \nabla^2 \varphi \, = \, 0 \, .
	\end{align}
Hence, $\varphi$ is a massless scalar field in AdS$_3$.
In order to construct bulk localized states of $\varphi$, we can use the discussion in Ref.\,\cite{Miyaji:2015fia}.
The massless scalar field leads to $(h, \bar{h})= (1,1)$ and this is the natural boundary condition from the point of view of the Maxwell theory in AdS$_3$ \cite{Marolf:2006nd}.
Since
	\begin{align}
		\mathcal{F}^I|_{\partial \mathcal{M}} \, \propto \, \epsilon^{IJ} D_J \varphi|_{\partial \mathcal{M}} \, ,
	\end{align}
and the asymptotic expansion of the massless scalar field is given by
	\begin{align}
		\varphi(x,z) \, = \, \varphi_{(0)}(x) \, + \, z^2 \varphi_{(2)}(x) \, + \, \cdots \, ,
	\end{align}
fixing $\mathcal{F}^I$ corresponds to fixing $\varphi_{(0)}(x)$.

Now let us explicitly construct the Maxwell bulk-local state.
First, we define $|\Phi(x) \rangle$ as the MNSTW \cite{Miyaji:2015fia} scalar bulk-local state constructed from the primary with $(h,\bar h)=(1,1)$.
Now the dual-field-strength and field-strength states are defined by
	\begin{align}
		|\widetilde{F}_\mu(x) \rangle \, := \, \nabla_\mu |\Phi(x) \rangle \, , \qquad |F_{\mu\nu}(x) \rangle \, := \, \epsilon_{\mu\nu}{}^\rho \nabla_\rho |\Phi(x) \rangle \, .
	\end{align}
The Maxwell equation follows from the symmetry of second derivatives of a scalar, while the Bianchi identity follows from the massless Klein--Gordon equation:
\begin{align}
\nabla^\mu|F_{\mu\nu}(x)\rangle&=0,
\qquad
\nabla_{[\rho}|F_{\mu\nu]}(x)\rangle=0.
\end{align}
Thus the construction describes the single local degree of freedom of the Maxwell field.
The corresponding gauge-invariant two-point function follows immediately as
	\begin{align}
		\langle F_{\mu\nu}(x) F_{\rho\sigma}(y) \rangle
        \, = \, \epsilon_{\mu\nu}{}^\alpha \epsilon_{\rho\sigma}{}^\beta \, \nabla^x_\alpha \, \nabla^y_\beta \, G_{\Delta=2}(x,y) \, ,
	\end{align}
where $G_{\Delta=2}(x,y)$ is the massless scalar propagator.

\subsubsection{The Maxwell state from the shortening limit}
\label{sec:spinone-limit-bridge}

The scalar construction above can be related directly to the massive vector states of Section~\ref{sec:ads}.
We take the massless limit of the normalized global descendants already introduced in eq.~\eqref{eq:ads-normalized-descendants}.
In this limit, the generator action separates a current module from a scalar module.
The lowest state in the scalar module is a vector descendant.
We then use this scalar sector to take the limit of the bulk local states themselves.
We work in the free one-particle sector.
Throughout this calculation, the AdS radius $\ell$ is fixed and
\begin{equation}
 M=m\ell=\Delta-1\longrightarrow0^+,
 \qquad H=\mathcal L_0+\overline{\mathcal L}_0.
 \label{eq:spinone-limit-parameters}
\end{equation}
This is a massless limit at fixed radius, not the flat limit considered in Section~\ref{sec:flat-limit}.
The physical energy at the center is $H/\ell$.
We leave the common vector normalization $\mathcal N_A$ arbitrary; all limits below explicitly separate this normalization from the descendant coefficients.

Consider the positive-helicity primary $|v_M^{(+)}\rangle=|h_+,\bar h_+\rangle_+$, normalized to unity, with
\begin{equation}
 (h_+,\bar h_+)=(1+M/2,M/2).
 \label{eq:spinone-limit-primary}
\end{equation}
Making the $M$ dependence of the basis in eq.~\eqref{eq:ads-normalized-descendants} explicit, we write
\begin{equation}
 |k,l;M\rangle_+
 =\frac{\mathcal L_{-1}^{k}\overline{\mathcal L}_{-1}^{l}}
        {\sqrt{k!\,l!\,(2+M)_k(M)_l}}|v_M^{(+)}\rangle,
 \qquad k,l\in\mathbb Z_{\ge0}.
 \label{eq:spinone-limit-normalized-basis}
\end{equation}
These states are orthonormal for every $M>0$.
When the additional label $M$ is omitted below, its dependence is understood as in eq.~\eqref{eq:spinone-limit-normalized-basis}.

The holomorphic generators act as
\begin{equation}
 \begin{aligned}
 \mathcal L_0|k,l;M\rangle_+
 &=\left(1+\frac M2+k\right)|k,l;M\rangle_+,\\
 \mathcal L_{-1}|k,l;M\rangle_+
 &=\sqrt{(k+1)(k+2+M)}\,|k+1,l;M\rangle_+,\\
 \mathcal L_1|k,l;M\rangle_+
 &=\sqrt{k(k+1+M)}\,|k-1,l;M\rangle_+,
 \end{aligned}
 \label{eq:spinone-limit-holomorphic-action}
\end{equation}
while the antiholomorphic generators act as
\begin{equation}
 \begin{aligned}
 \overline{\mathcal L}_0|k,l;M\rangle_+
 &=\left(\frac M2+l\right)|k,l;M\rangle_+,\\
 \overline{\mathcal L}_{-1}|k,l;M\rangle_+
 &=\sqrt{(l+1)(l+M)}\,|k,l+1;M\rangle_+,\\
 \overline{\mathcal L}_1|k,l;M\rangle_+
 &=\sqrt{l(l-1+M)}\,|k,l-1;M\rangle_+.
 \end{aligned}
 \label{eq:spinone-limit-lowering-matrix}
\end{equation}
All these matrix elements have finite limits.
In particular, the only matrix elements connecting the $l=0$ tower to the $l\ge1$ tower are
\begin{equation}
 \begin{aligned}
 \overline{\mathcal L}_{-1}|k,0;M\rangle_+
 &=\sqrt M\,|k,1;M\rangle_+,\\
 \overline{\mathcal L}_1|k,1;M\rangle_+
 &=\sqrt M\,|k,0;M\rangle_+.
 \end{aligned}
 \label{eq:spinone-limit-decoupling}
\end{equation}
Both vanish as $M\to0^+$.
The holomorphic generators preserve $l$, so the two towers become orthogonal invariant sectors in the massless limit.

On the $l=0$ tower, the antiholomorphic generators all act trivially in the limit, whereas the holomorphic generators give the weight-one highest-weight representation.
This sector is therefore the current module $D(1,0)$, which transforms in the same way as the chiral current, at least for the global modes.
For the $l\ge1$ tower, set $j=l-1$.
The antiholomorphic eigenvalue and ladder coefficients become
\begin{equation}
 \begin{aligned}
 l+\frac M2&\longrightarrow j+1,\\
 \sqrt{l(l-1+M)}&\longrightarrow\sqrt{j(j+1)},\\
 \sqrt{(l+1)(l+M)}&\longrightarrow\sqrt{(j+1)(j+2)}.
 \end{aligned}
 \label{eq:spinone-limit-scalar-matrix-elements}
\end{equation}
These are precisely the matrix elements of a weight-one antiholomorphic module, including the vanishing lowering coefficient at $j=0$.
The holomorphic coefficients in eq.~\eqref{eq:spinone-limit-holomorphic-action} likewise tend to those of a weight-one module.
Thus, the $l\ge1$ sector is $D(1,1)$, and
\begin{equation}
 D(1+M/2,M/2)\longrightarrow D(1,0)\oplus D(1,1).
 \label{eq:spinone-limit-module-split}
\end{equation}
Here, $D(h,\bar h)$ denotes the positive-energy global highest-weight module with the indicated weights.
Equation~\eqref{eq:spinone-limit-module-split} describes the limiting generator action on finite linear combinations of normalized descendants.
In contrast, setting $M=0$ in the unnormalized module and quotienting its null states retains only $D(1,0)$.
The two procedures retain different sectors and need not define the same physical massless limit.

For the negative-helicity branch, take the normalized primary $|v_M^{(-)}\rangle$ with weights $(h_-,\bar h_-)=(M/2,1+M/2)$.
The same argument applies with the two modules interchanged.
Now the matrix elements between $k=0$ and $k=1$ vanish as $\sqrt M$.
The $k=0$ tower gives $D(0,1)$, while the $k\ge1$ tower gives a scalar module after shifting the holomorphic level by one:
\begin{equation}
 D(M/2,1+M/2)\longrightarrow D(0,1)\oplus D(1,1).
 \label{eq:spinone-limit-negative-module-split}
\end{equation}
When both vector branches are present, the two scalar modules are separate orthogonal copies.
These statements concern the global modules used in the local-state construction, rather than a decomposition of the full Virasoro Hilbert space.

We now describe the $(1,1)$ sector explicitly.
In the positive branch, its lowest state is the limit of the normalized descendant $|0,1;M\rangle_+$.
At finite $M$, eqs.~\eqref{eq:spinone-limit-holomorphic-action} and \eqref{eq:spinone-limit-lowering-matrix} give
\begin{equation}
 \begin{aligned}
 \mathcal L_0|0,1;M\rangle_+
 &=\left(1+\frac M2\right)|0,1;M\rangle_+,\\
 \overline{\mathcal L}_0|0,1;M\rangle_+
 &=\left(1+\frac M2\right)|0,1;M\rangle_+,\\
 \mathcal L_1|0,1;M\rangle_+&=0,\\
 \overline{\mathcal L}_1|0,1;M\rangle_+
 &=\sqrt M\,|0,0;M\rangle_+.
 \end{aligned}
 \label{eq:spinone-limit-highest-weight}
\end{equation}
Its limit, denoted by $|0,0\rangle_{S,+}:= \lim_{M\to 0}|0,1;M\rangle_+$, has unit norm and weights $(1,1)$ and is annihilated by both lowering generators.
It is therefore a scalar primary, even though $|0,1;M\rangle_+$ is not a primary at finite $M$.
In terms of the original vector primary, the same descendant is
\begin{equation}
 |0,1;M\rangle_+
 =\frac{\overline{\mathcal L}_{-1}}{\sqrt M}|v_M^{(+)}\rangle.
 \label{eq:spinone-limit-scalar-seed}
\end{equation}
The factor $M^{-1/2}$ is exactly the normalization already present in eq.~\eqref{eq:spinone-limit-normalized-basis}, since
\begin{equation}
 \bigl\|\overline{\mathcal L}_{-1}|v_M^{(+)}\rangle\bigr\|^2
 =\langle v_M^{(+)}|
   [\overline{\mathcal L}_1,\overline{\mathcal L}_{-1}]
   |v_M^{(+)}\rangle=M.
 \label{eq:spinone-limit-null-norm}
\end{equation}
For the negative branch, the scalar primary $|0,0\rangle_{S,-}$ is similarly the limit of $|1,0;M\rangle_-=M^{-1/2}\mathcal L_{-1}|v_M^{(-)}\rangle$.

The normalization of the full scalar descendant tower follows directly from the same basis.
For $l\ge1$, using $(M)_l=M(1+M)_{l-1}$ in eq.~\eqref{eq:spinone-limit-normalized-basis} gives the exact finite-$M$ identity
\begin{equation}
 |k,l;M\rangle_+
 =\frac{\mathcal L_{-1}^{k}\overline{\mathcal L}_{-1}^{l-1}
           |0,1;M\rangle_+}
        {\sqrt{k!\,l!\,(2+M)_k(1+M)_{l-1}}}.
 \label{eq:spinone-limit-descendants-finite}
\end{equation}
At $M=0$, its denominator becomes the scalar normalization because $l!(1)_{l-1}=(l-1)!(2)_{l-1}$.
In each of the positive and negative branches, we can write the scalar descendant states as
\begin{equation}
 |k,j\rangle_{S,\lambda}
 =\frac{\mathcal L_{-1}^{k}\overline{\mathcal L}_{-1}^{j}}
        {\sqrt{k!\,j!\,(2)_k(2)_j}}|0,0\rangle_{S,\lambda},
 \qquad k,j\in\mathbb Z_{\ge0},\quad\lambda=\pm1,
 \label{eq:spinone-limit-scalar-descendants}
\end{equation}
with the basis maps
\begin{equation}
 \begin{aligned}
 |k,l;M\rangle_+&\longrightarrow|k,l-1\rangle_{S,+}
 &&(l\ge1),\\
 |k,l;M\rangle_-&\longrightarrow|k-1,l\rangle_{S,-}
 &&(k\ge1).
 \end{aligned}
 \label{eq:spinone-limit-basis-map}
\end{equation}
These scalar descendants satisfy
\begin{equation}
 \begin{aligned}
 H|k,j\rangle_{S,\lambda}
 &=(2+k+j)|k,j\rangle_{S,\lambda},\\
 (\mathcal L_0-\overline{\mathcal L}_0)|k,j\rangle_{S,\lambda}
 &=(k-j)|k,j\rangle_{S,\lambda}.
 \end{aligned}
 \label{eq:spinone-limit-scalar-spectrum}
\end{equation}
Thus, the primary has dimension two and spin zero, and the physical descendant energies are $(2+k+j)/\ell$.
This is the standard normalizable massless-scalar module used in the Maxwell construction above.
The label $\lambda$ records the parent vector branch, not a scalar helicity.

To determine the corresponding local state, we define one scalar center state in each limiting module by
\begin{equation}
 |\Phi_{S,\lambda}\rangle_0
 =\sum_{k=0}^{\infty}(-1)^k|k,k\rangle_{S,\lambda},
 \qquad \lambda=\pm1.
 \label{eq:spinone-limit-scalar-center}
\end{equation}
The subscript $0$ denotes the center of AdS, and the overall scalar normalization in this definition has been set to one.
As with the massive local states, this series defines a distributional state because its norm diverges.
All state limits in what follows hold after multiplication by $e^{-\varepsilon H}$ with any fixed $\varepsilon>0$.

From eq.~\eqref{eq:ads-coefficient-squares}, we obtain
\begin{equation}
 Ma_k^2\longrightarrow k(k+1),\qquad
 Mb_k^2\longrightarrow(k+1)^2,\qquad
 Mc_k^2\longrightarrow(k+1)(k+2).
 \label{eq:spinone-limit-coefficients}
\end{equation}
The first relation includes $k=0$, for which $Ma_0^2=M$.
Keeping the twist phases in eqs.~\eqref{eq:ads-state-plus}--\eqref{eq:ads-state-minus}, we define
\begin{equation}
 |\Psi_q^{(\lambda)}\rangle
 :=\lim_{M\to0}\frac{\sqrt M}{\mathcal N_A}
                  |\Phi_q^{(\lambda)}\rangle.
 \label{eq:spinone-limit-residue-definition}
\end{equation}
Notice that this normalization differs from the previous normalization of the descendants.
Here, we expand the bulk local state in the normalized basis, but the coefficients diverge as $M\to0$.
The overall factor $\sqrt{M}$ extracts the finite part of the expansion. Equations~\eqref{eq:spinone-limit-basis-map} and \eqref{eq:spinone-limit-coefficients} then yield
\begin{align}
 |\Psi_1^{(+)}\rangle
 &=\sum_{k=1}^{\infty}(-1)^k\sqrt{k(k+1)}|k,k-1\rangle_{S,+}
 =-\mathcal L_{-1}|\Phi_{S,+}\rangle_0,
 \label{eq:spinone-limit-residue-plus}\\
 |\Psi_0^{(+)}\rangle
 &=-\sqrt2\sum_{k=0}^{\infty}(-1)^k(k+1)|k,k\rangle_{S,+}
 =-\frac{H}{\sqrt2}|\Phi_{S,+}\rangle_0,
 \label{eq:spinone-limit-residue-zero}\\
 |\Psi_{-1}^{(+)}\rangle
 &=\sum_{k=0}^{\infty}(-1)^k\sqrt{(k+1)(k+2)}|k,k+1\rangle_{S,+}
 =\overline{\mathcal L}_{-1}|\Phi_{S,+}\rangle_0.
 \label{eq:spinone-limit-residue-minus}
\end{align}
For instance, the first operator identity follows by shifting the summation index in
\begin{equation}
 \mathcal L_{-1}|k,k\rangle_{S,+}
 =\sqrt{(k+1)(k+2)}|k+1,k\rangle_{S,+}.
 \label{eq:spinone-limit-scalar-ladder}
\end{equation}
The middle identity uses the scalar energy eigenvalue $2+2k$.
For the negative-helicity branch, the basis map is
$|k+1,l\rangle_-\to|k,l\rangle_{S,-}$.
Its phases in eqs.~\eqref{eq:ads-state-negative-minus}--\eqref{eq:ads-state-negative-plus} give
\begin{equation}
 \begin{split}
 |\Psi_1^{(-)}\rangle&=\mathcal L_{-1}|\Phi_{S,-}\rangle_0,\\
 |\Psi_0^{(-)}\rangle&=\frac{H}{\sqrt2}|\Phi_{S,-}\rangle_0,\\
 |\Psi_{-1}^{(-)}\rangle&=-\overline{\mathcal L}_{-1}|\Phi_{S,-}\rangle_0.
 \end{split}
 \label{eq:spinone-limit-negative-residues}
\end{equation}
Thus, in each branch, the different components of the vector state are obtained by acting with operators on the same scalar state.

These residues are precisely scalar-gradient multiplets.
To see this without choosing bulk coordinates, introduce the dimensionless generators complementary to the center stabilizer,
\begin{equation}
 Q_1=\frac{\mathcal L_{-1}-\overline{\mathcal L}_1}{\sqrt2},
 \qquad Q_0=H,
 \qquad Q_{-1}=\frac{\mathcal L_1-\overline{\mathcal L}_{-1}}{\sqrt2}.
 \label{eq:spinone-limit-displacements}
\end{equation}
The scalar center state obeys
$(\mathcal L_1+\overline{\mathcal L}_{-1})|\Phi_{S,\lambda}\rangle_0
 =(\mathcal L_{-1}+\overline{\mathcal L}_1)|\Phi_{S,\lambda}\rangle_0=0$.
Equations~\eqref{eq:spinone-limit-residue-plus}--\eqref{eq:spinone-limit-negative-residues} can therefore be written uniformly as
\begin{equation}
 |\Psi_q^{(\lambda)}\rangle
 =-\frac{\lambda}{\sqrt2}Q_q|\Phi_{S,\lambda}\rangle_0.
 \label{eq:spinone-limit-gradient}
\end{equation}
The operators $Q_q/\ell$ generate displacements of the center in a spherical tangent basis; the phases relating this basis to real coordinate derivatives are the same as in the vector local-state conditions.
In particular, one finds
\begin{equation}
 (\mathcal{L}_1+\overline{\mathcal{L}}_{-1})|\Psi_1^{(+)}\rangle
 =-2\mathcal L_0|\Phi_{S,+}\rangle_0
 =\sqrt2|\Psi_0^{(+)}\rangle,
 \qquad
 (\mathcal{L}_1+\overline{\mathcal{L}}_{-1})|\Psi_0^{(+)}\rangle=\sqrt{2}|\Psi_{-1}^{(+)}\rangle,
 \label{eq:spinone-limit-covariance}
\end{equation}
with the remaining conditions following from the same commutators.
This provides a component-by-component check that the limiting multiplet has the vector covariance required of the Maxwell dual field strength.

We can also confirm the relation by computing the inner product.
Let $\chi=D/\ell>0$ be the dimensionless separation used in the AdS two-point functions, and define
\begin{equation}
 \mathcal K_q(\chi)
 :=\lim_{M\to0}\frac{M}{|\mathcal N_A|^2}G_{qq}^{(+)}(\chi).
 \label{eq:spinone-limit-overlap-definition}
\end{equation}
From eqs.~\eqref{eq:ads-Gpp}--\eqref{eq:ads-Gmm}, we obtain
\begin{equation}
 \mathcal K_1=\mathcal K_{-1}=\frac1{4\sinh^3\chi},
 \qquad
 \mathcal K_0=\frac{\cosh\chi}{2\sinh^3\chi}.
 \label{eq:spinone-limit-overlap-result}
\end{equation}
The scalar two-point function is
\begin{equation}
 S(\chi):={}_0\langle\Phi_{S,\lambda}|e^{-\chi H}
                       |\Phi_{S,\lambda}\rangle_0
 =\sum_{k=0}^{\infty}e^{-(2+2k)\chi}
 =\frac{e^{-\chi}}{2\sinh\chi}
 =\frac{\coth\chi-1}{2}.
 \label{eq:spinone-limit-scalar-overlap}
\end{equation}
Differentiation with respect to $\chi$ gives
\begin{equation}
 \mathcal K_0=\frac12S''(\chi),
 \qquad
 \mathcal K_{\pm1}=-\frac{S'(\chi)}{2\sinh\chi}.
 \label{eq:spinone-limit-gradient-overlaps}
\end{equation}
Taking derivatives with respect to $\chi$ allows us to act with $H$ on the state.
Thus, $\mathcal K$ can be written in terms of derivatives of the scalar two-point function.
For $\mathcal K_1$, we use the commutation relations to obtain
\begin{equation}
    {}_0\langle\Phi_{S,\lambda}|Q_{-1}e^{-\chi H}Q_1|\Phi_{S,\lambda}\rangle_0=\frac{ {}_0\langle\Phi_{S,\lambda}|e^{-\chi H}H|\Phi_{S,\lambda}\rangle_0}{\sinh \chi},
\end{equation}
and
\begin{equation}
    {}_0\langle\Phi_{S,\lambda}|H e^{-\chi H}|\Phi_{S,\lambda}\rangle_0=-S'(\chi).
\end{equation}
This confirms eq.~\eqref{eq:spinone-limit-gradient-overlaps}, and the same argument applies to $\mathcal{K}_{-1}$.

To relate the massless limit to a physical field, we restore the radius in the first-order equation~\eqref{eq:ads-hkll-self-dual-equation}.
With the same choice of helicity and orientation, the dual field strength satisfies
\begin{equation}
 \widetilde F^{(\lambda)}_\mu
 :=\epsilon_\mu{}^{\nu\rho}\nabla_\nu A^{(\lambda)}_\rho
 =-\lambda m A^{(\lambda)}_\mu.
 \label{eq:spinone-limit-dual-field}
\end{equation}
Combining this equation with \eqref{eq:spinone-limit-gradient} gives, in the regulated sense stated above,
\begin{equation}
 |\widetilde F_q^{(\lambda)}\rangle
 =\frac{\sqrt M\,\mathcal N_A}{\sqrt2\,\ell}
       Q_q|\Phi_{S,\lambda}\rangle_0
       +\mathcal{O}\!\left(\sqrt M\,\mathcal N_A\right).
 \label{eq:spinone-limit-physical-strength}
\end{equation}
A nonzero finite field-strength limit therefore requires that
$\sqrt M\,\mathcal N_A$ approach a nonzero finite constant at fixed $\ell$.
Its value fixes the scalar normalization, or equivalently the Maxwell coupling in eq.~\eqref{eq:I_MW}.

%%%%%%%%%%%%%%%%%%%%%%%%%%%%%%%%%%%%%%%%%%%%%%%%%%%%%%%%
%%%%%%%%%%%%%%%%%%%%%%%%%%%%%%%%%%%%%%%%%%%%%%%%%%%%%%%%
\subsection{Chern-Simons theory}
\label{sec:chern-simons}
%%%%%%%%%%%%%%%%%%%%%%%%%%%%%%%%%%%%%%%%%%%%%%%%%%%%%%%%
In AdS$_3$/CFT$_2$, the usual holomorphic/anti-holomorphic $U(1)$ currents in CFT$_2$ are associated to the Chern-Simons theory in AdS$_3$ \cite{Kabat:2012hp, Terashima:2017gmc}.
Now we study the $U(1)$ Chern-Simons theory at level $k$ in AdS$_3$:
	\begin{align}
		I_{\textrm{CS}} \, = \, \frac{k}{4\pi} \int_{\mathcal{M}} A \wedge d A \, + \, \frac{k}{4\pi} \int_{\mathcal{M}} \bar{A} \wedge d \bar{A} \, .
	\end{align}
There is no nontrivial local gauge-invariant bulk wavefunction in this theory.
However, together with gravity, the asymptotic symmetry is now enhanced to the Virasoro + current algebras \cite{Keranen:2014ava,Zhao:2020qmn}.
This is explicitly written as
	\begin{align}
		&[\mathcal{L}_m, \mathcal{L}_n] \, = \, (m-n) \mathcal{L}_{m+n} \, + \, \frac{c}{12} \, m(m^2 - 1) \delta_{m+n,0} \, , \\
		&[\ol{\mathcal{L}}_m, \ol{\mathcal{L}}_n] \, = \, (m-n) \ol{\mathcal{L}}_{m+n} \, + \, \frac{\ol{c}}{12} \, m(m^2 - 1) \delta_{m+n,0} \, , \\
		&[\mathcal{L}_m, \ol{\mathcal{L}}_n] \, = \, 0 \, , \\[2pt]
		&[\mathcal{J}_m, \mathcal{J}_n] \, = \, k m \, \delta_{m+n,0} \, , \qquad [\ol{\mathcal{J}}_m, \ol{\mathcal{J}}_n] \, = \, k m \, \delta_{m+n,0} \, , \qquad [\mathcal{J}_m, \ol{\mathcal{J}}_n] \, = \, 0 \, , \\
		&[\mathcal{L}_m, \mathcal{J}_n] \, = \, - n \, \mathcal{J}_{m+n} \, , \ \qquad [\mathcal{L}_m, \ol{\mathcal{J}}_n] \, = \, 0 \, , \\
		&[\ol{\mathcal{L}}_m, \ol{\mathcal{J}}_n] \, = \, - n \, \ol{\mathcal{J}}_{m+n} \, , \qquad [\ol{\mathcal{L}}_m, \mathcal{J}_n] \, = \, 0 \, .
	\end{align}

Now if we define
	\begin{align}
		&L_n(\ell) \, := \, \mathcal{L}_n - (-1)^n \, \ol{\mathcal{L}}_{-n} \, , \qquad
		M_n(\ell) \, := \, \frac{1}{\ell} \Big( \mathcal{L}_n + (-1)^n \, \ol{\mathcal{L}}_{-n} \Big) \, , \\
		&J_n(\ell) \, := \, \mathcal{J}_n - (-1)^n \, \ol{\mathcal{J}}_{-n} \, , \qquad
		P_n(\ell) \, := \, \frac{1}{\ell} \Big( \mathcal{J}_n + (-1)^n \, \ol{\mathcal{J}}_{-n} \Big) \, ,
	\end{align}
then the algebras generated by the new generators are given by
	\begin{align}
		&[L_m(\ell), L_n(\ell)] \, = \, (m-n) L_{m+n}(\ell) \, , \\
		&[L_m(\ell), M_n(\ell)] \, = \, (m-n) M_{m+n}(\ell) \, + \, \frac{c_M}{12} \, m(m^2 - 1) \delta_{m+n,0} \, , \\
		&[M_m(\ell), M_n(\ell)] \, = \, \frac{(m-n)}{\ell^2} L_{m+n}(\ell) \, , \\[2pt]
		&[J_m(\ell), J_n(\ell)] \, = \, 0 \, , \qquad [P_m(\ell), P_n(\ell)] \, = \, 0 \, , \qquad [J_m(\ell), P_n(\ell)] \, = \, \frac{2k}{\ell} \, m \, \delta_{m+n,0} \, , \\
		&[L_m(\ell), J_n(\ell)] \, = \, - n \, J_{m+n}(\ell) \, , \ \qquad [L_m(\ell), P_n(\ell)] \, = \, - n \, P_{m+n}(\ell) \, , \\
		&[M_m(\ell), J_n(\ell)] \, = \, - n \, P_{m+n}(\ell) \, , \ \qquad [M_m(\ell), P_n(\ell)] \, = \, - \frac{n}{\ell^2} \, J_{m+n}(\ell) \, .
	\end{align}
Therefore, if we take the flat limit $\ell \to \infty$, we obtain the BMS$_3$ and two affine $U(1)$ current algebras discussed in Ref.\,\cite{Detournay:2016sfv} as
	\begin{align}
		&[L_m, L_n] \, = \, (m-n) L_{m+n} \, , \\
		&[L_m, M_n] \, = \, (m-n) M_{m+n} \, + \, \frac{c_M}{12} \, m(m^2 - 1) \delta_{m+n,0} \, , \\
		&[M_m, M_n] \, = \, 0 \, , \\[2pt]
		&[J_m, J_n] \, = \, 0 \, , \qquad [P_m, P_n] \, = \, 0 \, , \qquad [J_m, P_n] \, = \, \kappa_P \, m \, \delta_{m+n,0} \, , \\
		&[L_m, J_n] \, = \, - n \, J_{m+n} \, , \ \qquad [L_m, P_n] \, = \, - n \, P_{m+n} \, , \\
		&[M_m, J_n] \, = \, - n \, P_{m+n} \, , \ \qquad [M_m, P_n] \, = \, 0 \, ,
	\end{align}
where we defined
	\begin{align}
		\kappa_P \, = \, \lim_{\ell \to \infty} \frac{2k}{\ell} \, .
	\end{align}

\section{Conclusion and discussion}
\label{sec:conclusion}

We have constructed the bulk local states of a massive spin-1 field
in $\mathrm{AdS}_3/\mathrm{CFT}_2$ and in three-dimensional Carrollian
holography. We have also established their correspondence in the flat limit
within the free one-particle sector. By solving the covariant conditions, we obtain the three vector components in two branches of
definite helicity. Their inner products reproduce the vector
two-point functions. The AdS descendant expansion also agrees term
by term with the global HKLL reconstruction. The two branches combine
into the parity-invariant Proca field, while each branch separately
carries the single propagating mode of Maxwell-Chern-Simons theory in the massless case.

On the Carrollian side, we construct the states in the induced
representation. For a massive spin-1 field, the covariance and Casimir
conditions admit no nonzero solution in the BMS highest-weight modules
considered here. The AdS contraction preserves both mass and helicity.
Resumming the descendant expansion exactly, we obtain the Carrollian
states coefficient by coefficient in an ordered basis. With the
corresponding normalization of the states and algebraic duals, the
separated-point two-point functions also agree. This extends the
AdS-flat correspondence from the symmetry algebra to explicit local
states and their tensor correlators.

The bulk local states with the regulator also encode the geometry of a spatial slice.
Their Bures information metric has the hyperbolic form. We calculate
its coefficient exactly and find that it depends on the vector
component and the regulator. The phase of the overlap determines a
Berry connection proportional to the spatial spin connection. The
proportionality factor is fixed by the local rotation weight. At fixed
physical resolution, the component dependence of the metric remains
in the flat limit, including the nonrelativistic ratio $1:2:3$. In
contrast, the Berry curvature vanishes on fixed physical scales in the flat limit.

We also take the massless limit at fixed AdS radius and show how the
propagating Maxwell sector is contained in the massive construction.
In the normalized descendant basis, this limit separates each global
vector module into a chiral current module and a scalar module of
weights $(1,1)$. The appropriately rescaled residues of all three local
vector components are gradients of the scalar bulk local state.
Their overlaps reproduce derivatives of the scalar two-point function.
With the field normalization matched, we obtain a direct connection
to the Maxwell dual field strength. This normalized limit
differs from the null-state quotient that retains only the current
module. In the  Chern-Simons analysis, we identify the
boundary current algebra and its flat contraction. This distinguishes
the topological vector sector from the propagating Maxwell sector.

        There are several future directions concerning operator reconstruction and interactions.

        \noindent\textbf{HKLL reconstruction in flat spacetime.} Can the Carrollian bulk local states be obtained by applying a reconstructed bulk operator to the vacuum, as in our AdS calculation?
                A first step is to identify CCFT operators whose action on a vacuum realizes the induced states constructed here, an issue also raised for scalar reconstruction in ref.\,\cite{Hao:2025btl}.
                It remains unclear whether local boundary operators suffice or whether nonlocal operators are needed.
                With the appropriate boundary operators identified, one could seek a vector smearing kernel by expanding in bulk modes or using Green's functions.
                Acting on the vacuum, the reconstructed operator should reproduce our descendant coefficients and the two-point functions obtained with the algebraic duals.
                To establish an operator correspondence beyond these vacuum states, its matrix elements between excited states would also need to be compared with those of the bulk field.

            \noindent\textbf{Interactions.} Our construction describes free fields in the one-particle sector.
                In AdS, incorporating interactions into bulk local states requires superpositions of contributions from different conformal families \cite{Nakayama:2015mva}.
                For a vector field coupled to matter, a concrete next step is to construct the first perturbative correction to the local state, including contributions from states containing several particles.
                The coefficients must be fixed by the interacting bulk equations and boundary correlation functions, since covariance alone does not specify the interaction.
                Matching a bulk three-point function would test the reconstruction beyond the free propagator.
                We could then take the flat limit of this correction and test the AdS--Carrollian correspondence at the same perturbative order.

    \acknowledgments
        We are grateful to Wu-Zhong Guo and Jiang Long for fruitful discussion.

        The work of KS is supported by JSPS KAKENHI Grant No.~23K13105.
        Y.S. is supported by special post-doctoral researchers program at RIKEN.
        Research at Perimeter Institute is supported in part by the Government of Canada through the Department of Innovation, Science and Economic Development and by the Province of Ontario through the Ministry of Colleges and Universities. PH is supported by the funding from HIMIS.
        We acknowledge OpenAI’s ChatGPT and DeepSeek for assistance with calculations and literature searches.
        All AI-assisted calculations and references were independently reviewed and verified by the authors, who take full responsibility for the content of the manuscript.

    \appendix

    \section{Proof of eq.\,\eqref{ea:adscft-to-flat-descendant}}
        \label{appendix:ads-descendant-conversion}

        We prove eq.\,\eqref{ea:adscft-to-flat-descendant} for arbitrary $p,q\in\mathbb{Z}_{\geq0}$.
        The finite-radius generators in eq.\,\eqref{eq: generator relation between flat and AdS} obey
        \begin{align}
            L_{-1}(\ell)=\mathcal{L}_{-1}+\overline{\mathcal{L}}_{1},\qquad
            L_{1}(\ell)=\mathcal{L}_{1}+\overline{\mathcal{L}}_{-1}.
            \label{eq:appendix-global-generators}
        \end{align}
        The highest weight conditions imply
        \begin{align}
            \mathcal{L}_{-1}^{p}|h,\overline{h}\rangle
            =L_{-1}(\ell)^{p}|h,\overline{h}\rangle.
            \label{eq:appendix-holomorphic-descendant}
        \end{align}
        Hence the left-hand side of eq.\,\eqref{ea:adscft-to-flat-descendant} can be written as
        \begin{align}
            \mathcal{L}_{-1}^{p}\overline{\mathcal{L}}_{-1}^{q}|h,\overline{h}\rangle
            =\overline{\mathcal{L}}_{-1}^{q}L_{-1}(\ell)^{p}|h,\overline{h}\rangle.
            \label{eq:appendix-starting-descendant}
        \end{align}

        To eliminate one antiholomorphic lowering generator, use
        $\overline{\mathcal{L}}_{-1}=L_{1}(\ell)-\mathcal{L}_{1}$.
        The three operators $\overline{\mathcal{L}}_{-1}$, $L_{1}(\ell)$, and $\mathcal{L}_{1}$ commute pairwise.
        For $n\geq1$, $m\geq1$, and $c\geq0$, the global conformal algebra then fixes the recursion to
        \begin{align}
            &\overline{\mathcal{L}}_{-1}^{n}L_{1}(\ell)^{c}L_{-1}(\ell)^{m}|h,\overline{h}\rangle\nonumber\\
            &\quad=
            \overline{\mathcal{L}}_{-1}^{n-1}L_{1}(\ell)^{c+1}L_{-1}(\ell)^{m}|h,\overline{h}\rangle\nonumber\\
            &\qquad
            -m(2h+m-1)\overline{\mathcal{L}}_{-1}^{n-1}
            L_{1}(\ell)^{c}L_{-1}(\ell)^{m-1}|h,\overline{h}\rangle.
            \label{eq:appendix-descendant-recursion}
        \end{align}
        When $m=0$, only the first term on the right-hand side is present.
        The coefficient in the second term follows from
        \begin{align}
            \mathcal{L}_{1}L_{-1}(\ell)^{m}|h,\overline{h}\rangle
            =m(2h+m-1)L_{-1}(\ell)^{m-1}|h,\overline{h}\rangle.
            \label{eq:appendix-sl2-action}
        \end{align}

        Apply eq.\,\eqref{eq:appendix-descendant-recursion} $q$ times to eq.\,\eqref{eq:appendix-starting-descendant}.
        Suppose the first term of the recursion is selected $b$ times and the second term is selected $q-b$ times.
        The resulting state is
        \begin{align}
            L_{1}(\ell)^{b}L_{-1}(\ell)^{p-q+b}|h,\overline{h}\rangle.
            \label{eq:appendix-terminal-descendant}
        \end{align}
        It exists only when $q-b\leq p$, which fixes the lower bound $b\geq\max(0,q-p)$ in eq.\,\eqref{ea:adscft-to-flat-descendant}.

        Each use of the second term lowers the power of $L_{-1}(\ell)$ by one.
        Its accumulated weight is therefore independent of the positions of those choices and equals
        \begin{align}
            (-1)^{q-b}\prod_{j=p-q+b+1}^{p}j(2h+j-1).
            \label{eq:appendix-path-weight}
        \end{align}
        There are $\binom{q}{b}$ ways to place the $b$ first-term choices among the $q$ recursion steps.
        The coefficient of eq.\,\eqref{eq:appendix-terminal-descendant} is consequently the product form of $C_{p,q}(b)$ in eq.\,\eqref{eq:ads-descendant-conversion-coefficient}.
        The gamma-function representation of each finite product, with the pole prescription stated below eq.\,\eqref{eq:ads-descendant-conversion-coefficient}, reproduces the second line of that equation and completes the proof.

        For a scalar primary with $h=\overline{h}$, the specialization $p=q=k$ is equivalent to the diagonal descendant identity in Ref.\,\cite{Hao:2025btl}, which is written with the opposite normal ordering.
        The first nontrivial case is
        \begin{align}
            \mathcal{L}_{-1}\overline{\mathcal{L}}_{-1}|h,\overline{h}\rangle
            =\left[L_{1}(\ell)L_{-1}(\ell)-2h\right]|h,\overline{h}\rangle,
        \end{align}
        which also follows directly from eq.\,\eqref{eq:appendix-global-generators} and the highest weight conditions.

    \bibliographystyle{JHEP}
    \bibliography{main.bib}

\end{document}